\documentclass[12pt]{article}

\usepackage[utf8]{inputenc}
\usepackage[margin=25truemm]{geometry}
\usepackage{amsthm}
\usepackage{ascmac}
\usepackage{latexsym}
\usepackage{indentfirst}
\usepackage{cancel}
\usepackage{comment}

\usepackage{amsmath,amssymb}
\usepackage{mathtools}
\usepackage{physics}
\usepackage{bbm}

\usepackage{graphicx}
\usepackage{hyperref}

\usepackage{cite}
\makeatletter
  \renewcommand{\theequation}
  {\arabic{section}.\arabic{equation}}
  \@addtoreset{equation}{section}
 \makeatother
\numberwithin{equation}{section}

\allowdisplaybreaks[1]

\begin{document}

\begin{titlepage}
\renewcommand{\thefootnote}{\fnsymbol{footnote}}
\begin{normalsize}
\begin{flushright}
\begin{tabular}{l}
Octorber 2024
\end{tabular}
\end{flushright}
\end{normalsize}

~~\\

\vspace*{0cm}
    \begin{Large}
       \begin{center}
         {Regularization of matrices 
         in the covariant derivative interpretation \\
         of matrix models  }
       \end{center}
    \end{Large}
\vspace{1cm}

\begin{center}
           Keiichiro H{\sc attori}$^{1)}$\footnote
            {
e-mail address:
hattori.keiichiro.17@shizuoka.ac.jp},
           Yuki M{\sc izuno}$^{1)}$\footnote
            {
e-mail address:
mizuno.yuhki.15@shizuoka.ac.jp}
           {and}
           Asato T{\sc suchiya}$^{1),2)}$\footnote
           {
e-mail address: tsuchiya.asato@shizuoka.ac.jp}
\\
      \vspace{1cm}

$^{1)}$
 {\it Graduate School of Science and Technology, Shizuoka University}\\
               {\it 836 Ohya, Suruga-ku, Shizuoka 422-8529, Japan}\\
        \vspace{0.3cm}

 $^{2)}$
{\it Department of Physics, Shizuoka University}\\
                {\it 836 Ohya, Suruga-ku, Shizuoka 422-8529, Japan}\\

\end{center}

\vspace{2cm}

\begin{abstract}
\noindent
We study regularization of matrices in the covariant derivative
interpretation of matrix models, a typical example of 
which is the type IIB matrix model.
The covariant derivative interpretation provides a possible way 
in which curved spacetimes are described by matrices, 
which are viewed as differential operators.
One needs to regularize the operators as matrices with 
finite size in order to apply the interpretation to nonperturbative
calculations such as numerical simulations.
We develop a regularization of the covariant derivatives 
in two dimensions by using the Berezin-Toeplitz quantization. 
As examples, we examine the cases of $S^2$ and $T^2$ in details.

\end{abstract}
\end{titlepage}
\vfil\eject

\setcounter{footnote}{0}

\tableofcontents

\section{Introduction} \label{sec:introduction}
Matrix models are expected to give a 
nonperturbative formulation of superstring theory \cite{Banks:1996vh,Ishibashi:1996xs,Dijkgraaf:1997vv,Berenstein:2002jq}.
One of those is the type IIB matrix 
model \cite{Ishibashi:1996xs} (for recent studies of the model, see \cite{Anagnostopoulos:2022dak,Hirasawa:2024dht,Asano:2024def,Asano:2024edo,Steinacker:2023cuf,Kumar:2023bxg,Steinacker:2023ntw,Steinacker:2024unq,Steinacker:2024huv,Brahma:2022ik,Laliberte:2023bai,Brandenberger:2024ddi,Laliberte:2024iof,Klinkhamer:2022frp,Hartnoll:2024csr,Kumar:2022giw}).
The action of the model is 
given by the dimensional reduction 
of $\mathcal{N}=1$ super Yang-Mills theory
in ten dimensions to zero dimension.
Thus, the spacetime does not exist in the definition of the model, 
but it 
emerges from the degrees of freedom of matrices. 
Also, string theory includes gravity, so does the model.
Thus, curved spacetimes should be described in terms of matrices in the model.

One of proposals for such description is 
the covariant derivative interpretation developed in
\cite{Hanada:2005vr} (see also \cite{Hanada:2006gg,Hanada:2006ei,Furuta:2006kk,Saitou:2007sf,Matsuo:2008yd,Asano:2012mn,Sakai:2017dxi,Sakai:2019cmj}, and see \cite{Steinacker:2024unq} and references
therein for another approach to this issue).
Its interesting features
are that the Einstein equations arise from the equations of motion of the model
and that the general coordinate transformation and the local Lorentz 
transformation are included in the U($N$) transformation of matrices.
In this interpretation, the matrices are viewed as
differential operators that can be 
expanded in terms of the covariant derivatives, which
possess the information of geometry.
However, the size of matrices in this interpretation is infinite.
It is necessary to make the size of matrices finite
in order to apply the interpretation to nonperturbative calculations 
such as numerical simulations (see \cite{Anagnostopoulos:2022dak,Hirasawa:2024dht} 
for recent progress in the numerical simulation of the type IIB matrix model).

In this paper, we study regularization of matrices 
in the covariant derivative 
interpretation of the matrix models.
While we consider the type IIB matrix model concretely, 
our results in this paper can be applied to the other models.
As a first step, we consider the two-dimensional case.
We regularize the covariant derivatives in two dimensions as matrices with finite size by using
the Berezin-Toeplitz (BT) 
quantization \cite{Bordemann:1993zv,Hawkins_2000}.
As examples, we examine the cases of $S^2$ and $T^2$
in details. For simplicity, we restrict ourselves 
to manifolds with the Euclidean signature
keeping the Euclidean version of the type IIB matrix model in mind,
while recent progress in the numerical simulations 
of the type IIB matrix model suggests
that the Lorentzian version is more relevant 
for describing the real world.

This paper is organized as follows. In section 2, we briefly review
the covariant derivative interpretation of matrix models 
and the BT quantization. 
In section 3, we devlelop a regularization of the covariant derivatives in two-dimensions 
by using the BT quantization. As concrete examples, 
we examine the case of $S^2$ and $T^2$ in details.
Section 4 is devoted to conclusion and discussion.
In appendices, some details of calculations are gathered.

\section{Reviews}

\subsection{Covariant derivative interpretation of matrix models} \label{sec:covariant_derivative_interpretation}
In this subsection, we give a brief review of the covariant derivative
interpretation of matrix models, which was developed in \cite{Hanada:2005vr}.

There arise two problems when matrices are naively interpreted
as covariant derivatives, namely $A_a=i\nabla_a$ where
$a$ is the index of the local Lorentz coordinates.
One is that covariant derivatives are not defined globally but 
transformed between coordinate patches.
The other is that in such interpretation
the product of two matrices $A_1$ and $A_2$ is calculated as
    \begin{equation}
      A_1 A_2 = i \partial_1 A_2 +i\Omega_1^{\;2c}A_c \ ,
    \label{product}
    \end{equation}
where $\Omega_a^{\;bc}$ is the spin connection.
The existence of the second term in \eqref{product}
indicates that the space on which $A_2$ acts is different from the one on which $A_1$ acts.
Hence, $A_1A_2$ cannot be regarded as the product of matrices.
These problems are overcome by
considering 
a fiber bundle with the representation 
space of the regular representation
of Spin($d$) being
the fiber, as seen in the following.

Let $M$ be a $d$-dimensional manifold and $\{ U_i\}$ be
a local coordinate system of $M$.
Suppose that 
$E_{prin}$ is a principal bundle over $M$ with
the structure group $G$ and 
$E_{reg}$ is 
a fiber bundle over $M$ with the representation space of 
the regular representation of $G$ denoted by $V_{reg}$
being the fiber.
Here $V_{reg}$ is the set of functions from $G$ to 
$\mathbb{C}$:
    \begin{equation}
      V_{reg} = \qty{f:G\to \mathbb{C}} \ .
    \end{equation}
The action of $h\in G$ on $f\in V_{reg}$ is defined by
    \begin{equation}
      \hat{h}f(g) = f(h^{-1}g) \ .
    \end{equation}
The regular representation is a reducible representation.
The Peter-Weyl theorem states that
its irreducible decomposition is given by
    \begin{equation}
        f(g) = \sum_{r:irr} c^{\langle r \rangle}_{ij} \sqrt{d_r} R^{\langle r\rangle}_{ij}(g) \ ,
        \label{Peter-Weyl}
    \end{equation}
where $r$ labels the irreducible representations, 
$d_r$ stands for the dimension of the $r$ representation, 
and $R^{\langle r\rangle}_{ij}(g)$ are elements of
the representation matrix of the $r$ representation for $g$.
Then, the index $i$ of $c^{\langle r\rangle}_{ij}$ is transformed as the 
dual representation of the $r$ representation under the action of $G$, while the index $j$ is invariant. The regular representation is,
therefore, decomposed as
    \begin{equation}
        V_{reg} = \underset{r}{\oplus} \underbrace{(V_r \oplus \dots \oplus V_r)}_{d_r} \ . 
    \end{equation}
In particular, for $G=$Spin(2), this decomposition
is nothing but the Fourier expansion.
We denote the set of global sections on $E_{reg}$ by
$\Gamma(E_{reg})$. $C^{\infty}(E_{prin})$ and $\Gamma(E_{reg})$ are
isomorphic:
    \begin{equation}
      C^{\infty}(E_{prin}) \simeq \Gamma (E_{reg}) \ .
    \end{equation}
Indeed, an element $f$ of $C^{\infty}(E_{prin})$ and an element $\tilde{f}$ of $\Gamma (E_{reg})$ 
are local functions from 
$U_i \times G$ to $\mathbb{C}$ that must satisfy
the same gluing condition on the overlap between $U_i$ and $U_j$:
for $x \in U_i \cup U_j$, 
    \begin{equation}
      f(x^{[i]},g) = f(x^{[j]},t_{ij}(x)g) \ ,
    \end{equation}
where $t_{ij}(x)$ is a transformation function.

Furthermore, there is the following isomorphism for an arbitrary
representation $r$ of $G$:
    \begin{equation} \label{eq:r_times_reg_and_directsum_of_regs}
      V_r \otimes V_{reg} \simeq \underbrace{V_{reg} \oplus \dots \oplus V_{reg}}_{d_r} \ ,
    \end{equation}
where $V_r$ is the representation space of the $r$ representation.
The isomorphism is constructed as follows.
Let $\Phi^i(g)$ be an element of $V_r \otimes V_{reg}$.
Namely, $h \in G$ acts on $\Phi^i(g)$ as 
    \begin{equation} \label{eq:r_times_reg}
      \hat{h}\Phi^i (g) = R^i{}_j(h) \Phi^j(h^{-1}g) \ ,
    \end{equation}
where $R^i{}_j(g)$ are the matrix elements of the $r$ representation
for $g$.  $\Phi^{(i)} (g)$ are defined by
\begin{equation} \label{isomorphism for regular rep}
  \Phi^{(i)} (g) = R^{(i)}{}_j(g^{-1}) \Phi^j(g)\ . 
\end{equation}
Here, while $R^{(i)}{}_j$ are the same quantities as $R^i{}_j$
that appear in \eqref{eq:r_times_reg},
the index $(i)$ is not transformed under the action of $G$ as shown 
below.
Acting $h \in G$ on $\Phi^{(i)}(g)$ leads to
    \begin{align}
    \hat{h}\Phi^{(i)}(g) &=
      R^{(i)}{}_j(g^{-1})R^j{}_k(h)\Phi^k(h^{-1}g)  \nonumber\\ 
      &= R^{(i)}{}_k((h^{-1}g)^{-1})\Phi^{k} (h^{-1}g) \nonumber\\
      &=\Phi^{(i)}(h^{-1}g) \ .
    \end{align}
The last equality indicates that 
each of $d_r$ elements of $\Phi^{(i)}(g)$ 
gives a regular representation of $G$.
As for the sections of $E_{reg}$, one can show in a similar way that
    \begin{equation} 
      \Gamma(E_r \otimes E_{reg}) \simeq \underbrace{\Gamma(E_{reg}) \oplus \dots \oplus \Gamma(E_{reg})}_{d_r} \ ,
      \label{isomorphism}
    \end{equation}
where $E_r$ is the fiber bundle with the representation space of
the $r$ representation
of $G$ being the fiber.

In what follows, we consider the case in which $G=$Spin($d$) or 
$G=$Spin${}_C$($d$).
The isomorphism (\ref{isomorphism}) enables to treat tensor fields 
as a direct sum of fields obeying the regular representation.
We will indeed see below that one can
regard the covariant derivative as an operator acting on $\Gamma(E_{reg})$.
The covariant derivative $\nabla_a$ is defined by \footnote{In the case of Spin${}_C$($d$), the U(1) part is added to the RHS of \eqref{covariant derivative}.}
    \begin{equation}
      \nabla_a = e^\mu_a \left( \partial_\mu + \frac{1}{2}\Omega_\mu{}^{bc}O_{bc} \right) \ ,
    \label{covariant derivative}
    \end{equation}
where $\Omega_\mu{}^{ab}$ are the spin connection and
$O_{bc}$ are generators of Spin($d$) whose action on $\Gamma(E_{reg})$ is given by
\begin{equation}
i \epsilon^{ab}(O_{ab}f)(x,g) = f(x,(1+\epsilon^{ab}M_{ab})^{-1}g) - f(x,g)
\end{equation}
with $M_{ab}$ being the representation matrices for the fundamental representation.
    
The covariant derivative $\nabla_a$ acts on $\Gamma(E_{reg})$ as
    \begin{equation}
      \nabla_a : \Gamma(E_{reg}) \to \Gamma(TM \otimes E_{reg}) 
      \  ,
    \end{equation}
where $TM$ is the tangent bundle. It is seen from the above argument that
$\nabla_{(a)}$ defined by
    \begin{equation}
      \nabla_{(a)} = R_{(a)}{}^b(g^{-1}) \nabla_b\ ,
    \end{equation}
act on $\Gamma(E_{reg})$ as
    \begin{equation}
      \nabla_{(a)} : \Gamma(E_{reg}) \to \underbrace{\Gamma(E_{reg}) \oplus \dots \oplus \Gamma(E_{reg})}_{d} \ .
    \end{equation}
The index $(a)$ is not that of a vector but the label of $d$ copies
of the regular representation.
Namely, each element of $\nabla_{(a)}$ is 
an endomorphism from $\Gamma(E_{reg})$ to $\Gamma(E_{reg})$.
Thus, the problem of the product in the covariant derivative
interpretation is resolved.

Next, let us see whether $\nabla_{(a)}$ is 
defined globally.
$\nabla^{[i]}_a$ and $\nabla^{[j]}_a$, 
which are defined in local patches $U_i$ and $U_j$, 
respectively, are related at a point in the overlap
between $U_i$ and $U_j$,
$x\in U_i \cup U_j$, as 
    \begin{equation}
      \nabla^{[i]}_a= R_a{}^b(t_{ij}(x))\nabla^{[j]}_b  \ .
    \end{equation}
By using this relation, $\nabla^{[i]}_{(a)}$ is calculated as
    \begin{align}
      \nabla^{[i]}_{(a)} &= R_{(a)}{}^b(g_{[i]}^{-1})\nabla^{[i]}_b 
      \nonumber\\
      &= R_{(a)}{}^b(g_{[i]}^{-1}) R_b{}^c(t_{ij}(x))\nabla^{[j]}_c 
      \nonumber\\
      &= R_{(a)}{}^b((t_{ij}^{-1}(x)g_{[i]})^{-1}) \nabla^{[j]}_b
      \nonumber \\
      &= R_{(a)}{}^b(g_{[j]}^{-1})\nabla^{[j]}_b  \nonumber\\
      &= \nabla^{[j]}_{(a)} \ ,
    \end{align}
which indeed indicates that $\nabla^{[i]}_{(a)}$ is defined globally.
In this manner, the matrices can be interpreted as the covariant 
derivatives:$A_{(a)}=i\nabla_{(a)}$.
In general, $A_{(a)}$ are viewed as $d$ endomorphisms on
$\Gamma (E_{reg})$ including fluctuations around a background 
$\nabla_{(a)}$ and expanded in $\nabla_{(a)}$ and $O_{ab}$.

One of the features of the covariant derivative interpretation is 
that the Einstein equations are derived from
the equations of motion of the matrix model.
The bosonic part of the action of the Euclidean type IIB matrix model
with a mass term
takes the form
    \begin{equation}
      S = -\frac{1}{4g^2}\mathrm{Tr}([A_{(a)},A_{(b)}]
      [A^{(a)},A^{(b)}]) + \frac{m^2}{g^2} \mathrm{Tr}(A_{(a)}A^{(a)}) 
      \ ,
    \end{equation}
where $a$ and $b$ run from 1 to 10.
The equations of motion derived from this action are 
    \begin{equation}
      [A^{(a)},[A_{(a)},A_{(b)}]] -2m^2 A_{(b)}= 0  \ .
    \end{equation}
These equations are equivalent to
    \begin{equation} \label{eq:eom_hkk}
      [A^a,[A_a,A_b]] -2m^2 A_b= 0 \ .
    \end{equation}
For simplicity, the following form of solutions is assumed:    
    \begin{equation}
    \begin{array}{ll}
      A_a = i \nabla_a  &  \mbox{for} \;\; a = 1, \ldots, d \ ,\\
      A_a = 0           &  \mbox{for} \;\; a = d+1,\ldots, 10 \ ,
    \end{array}
    \label{assumption of solution}
    \end{equation}
where $\nabla_a$ are defined in (\ref{covariant derivative}).
Substituting this into \eqref{eq:eom_hkk} yields
    \begin{equation}
      [\nabla^a,[\nabla_a,\nabla_b]] =  (\nabla^a R_{ab}{}^{cf})O_{cf} - R_b{}^c \nabla_c  \ ,
    \end{equation}
where $a, \ b,\ c,\ f=1, \ldots, d$.
Thus, \eqref{eq:eom_hkk} reduces to 
    \begin{equation}
      \nabla^a R_{ab}{}^{cf} = 0 \ ,\; \; 
      R_b{}^c = -2m^2 \delta_{bc} \ .
    \end{equation}
The first equations are obtained from 
the Bianchi identity in $d$ dimensions and the second
equations. The second equations are the Einstein equation
in $d$ dimensions with the 
cosmological constant
    \begin{equation}
      \Lambda = - (d-2)m^2 \ .
    \end{equation}

\subsection{Examples: $S^2$ and $T^2$} \label{subsec:hkk_S2}
In this subsection, as examples, 
we construct $\nabla_{(a)}$ for $S^2$ and $T^2$
as in \cite{Hanada:2005vr}.
The spin $s$ representation of $G=$Spin(2) is given by
\begin{equation}
  R^{(s)}(\theta) = e^{2is\theta} \ , 
  \label{spin s representation}
\end{equation}
where $\theta \in [0,2\pi)$ and $s$ is a half-integer or an integer. 
The action of the generators $O_{+-}$ on $\Gamma(E_{reg})$
is given by
\begin{equation}
  O_{+-} = \frac{1}{4i} \partial_\theta  \ ,
\end{equation}
where the suffices $\pm$ are defined through the equation
$e^{\pm} =e^1 \pm ie^2$ for a contravariant vector $e^a$.
    
We consider a unit $S^2$ embedded in $\mathbb{R}^3$.
$X^A$ ($A=1,2,3$) denote the embedding coordinates
and $(z,\bar{z})$ denotes the coordinates of the stereographic 
projection of $S^2$ from
the north pole to the $X^1-X^2$ plane.
They are related as
\begin{equation}
    X^1 = \frac{z+\bar{z}}{1+|z|^2} \ ,\;\; X^2 = -i\frac{z-\bar{z}}{1+|z|^2} \ ,\;\; X^3 = \frac{1-|z|^2}{1+|z|^2} \ .
    \label{embedding coordinates}
\end{equation}
The induced metric is represented in terms of $(z,\bar{z})$ as
\begin{equation}
  g_{z\bar{z}} =g_{\bar{z}z} = \frac{2}{(1+|z|^2)^2} \ , \;\; g_{zz}=g_{\bar{z}\bar{z}}=0 \ .
  \label{metric}
\end{equation}
We adopt the following zweibein that yields this metric:
\begin{equation}
    e^1 = \frac{dz + d\bar{z}}{1+|z|^2} \ , \; \; e^2 = -i \frac{dz - d\bar{z}}{1+|z|^2} \ .
    \label{zweibein}
\end{equation}
The inverse of the zweibein is given by
\begin{align}
e_1 = \frac{1}{2} (1+|z|^2) ( \partial_z + \partial_{\bar{z}}) \
      , \; \; e_2 = \frac{i}{2} (1+|z|^2) ( \partial_z - 
      \partial_{\bar{z}}) \ .
      \label{inverse of zweibein}
\end{align}
The spin connection is determined from the torsionless condition 
$de^a+\Omega^a{}_b\wedge e^b=0$ as
\begin{equation}
 \Omega^1{}_2 = -i \frac{\bar{z}dz - z d\bar{z}}{1+|z|^2} \ .
 \label{spin connection}
\end{equation}
$\nabla_{(a)}$ are calculated as
\begin{align}
&\nabla_{(+)} = \frac{1}{2}e^{-2i\theta}\qty( (1+|z|^2)\partial_z + \frac{i}{2}\bar{z}\partial_\theta) \ , \notag\\
&\nabla_{(-)} = \frac{1}{2}e^{2i\theta}\qty( (1+|z|^2)\partial_{\bar{z}} - \frac{i}{2}z\partial_\theta) \ .
\end{align}
It is easy to show that the following commutation relations hold:
\begin{align}
&[\nabla_{(+)},\nabla_{(-)}] = \frac{1}{4i} \partial_\theta = O_{+-} \ ,
\label{eq:nabla+_nabla-_commutator} \\
& [O_{+-},\nabla_{(\pm)}] = \mp \frac{1}{2} \nabla_{(\pm)} \ .
\label{eq:lorentz_nabla_commutator}
\end{align}
Thus, $\nabla_{(\pm)}$ and $O_{+-}$ form the su(2) Lie algebra.
This reflects the fact that $S^3\simeq \mathrm{SU}(2)$ is a Spin(2) principal bundle over $S^2$.

Next, we consider a torus $T^2$.
$T^2$ is defined by periodicity $x^a \sim x^a+2\pi$.
To cover all regions of $T^2$, we need four coordinate patches.
We denote them by $(z_i,\bar{z}_i)$ $(i=1,\dots,4)$, where $z_i=x^{[i]1}+ix^{[i]2}$.
The metric is given by
\begin{align}
    g_{z_i\bar{z}_i}=g_{\bar{z}_iz_i}=\frac{1}{2}\ , \;\; g_{z_iz_i}=g_{z_iz_i}=0 \ .
\end{align}
The spin connection for $T^2$ vanishes.
The covariant derivatives are calculated as
\begin{align}
    &\nabla_{(+)}^{[z_i]} = \frac{1}{2} e^{-2i\theta_i}\partial_{z_i} 
    \ ,\notag\\
    &\nabla_{(-)}^{[z_i]} = \frac{1}{2} e^{2i\theta_i}\partial_{\bar{z}_i} \ .
\end{align}

\subsection{Berezin-Toeplitz quantization} \label{subsec:BT_quantization}
In this subsection, we give a brief review of the BT quantization.
The BT quantization is a method for regularizing
fields on a manifold as matrices with finite size.
The matrices obtained by the BT quantization are called the Toeplitz 
operators. It is known that the Toeplitz operators satisfy
appropriate properties for regularization 
in the limit in which the matrix size goes to infinity.  
Here we review the BT quatization for a general fiber bundle 
over a closed Riemann surface \cite{Adachi:2021aux}.

Let $M$ be a closed Riemann surface with a Riemann metric $g$.
$\omega$ is the invariant volume form defined in terms of $g$ such that
the volume $V$ of $M$ is given by $V=\frac{1}{2\pi}\int_M \omega$.
Suppose that $E$ and $E'$ are fiber bundles over $M$ that possess the Hermite
inner product and the Hermite connection.
$\mathrm{Hom}(E,E')$ is the homomorphism bundle on $M$ whose fiber at $p \in M$ is a set of all linear maps from the fiber of $E$ at $p$ to that of $E'$ at $p$.
$\Gamma(E)$ denotes the set of all sections of $E$.
In the BT quantization, elements $\chi \in \Gamma(\mathrm{Hom}(E,E'))$ are mapped to matrices with finite size.

In order to perform the BT qunatization, one first considers
$\Gamma(S \otimes L^{\otimes N} \otimes E)$ and $\Gamma(S \otimes L^{\otimes N} \otimes E')$,
where $S$ is the spinor bundle over $M$, $N$ is a positive integer and
$L$ is a complex line bundle with a connection 1-form $A$ that satisfies $F=dA=\omega/V$.
This normalization implies that the 1st Chern number of $L$ is $\frac{1}{2\pi}\int_M F=1$. 
Then, $\chi \in \mathrm{Hom}(E,E')$ acts on $\psi \in \Gamma(S\otimes L^{\otimes N}\otimes E)$ as $\psi \to \chi \psi \in \Gamma(S\otimes L^{\otimes N }\otimes E')$. The Hermite inner products in the fibers of $S$, $L$ and $E$ induce an
inner product in $\Gamma(S\times L^{\otimes N}\times E)$ defined by
    \begin{equation} \label{eq:inner_product}
      (\psi,\psi') = \int_M \omega (\psi)^{\dagger} \cdot \psi'  \ ,
    \end{equation}
where $\psi ,\, \psi' \in \Gamma(S\otimes L^{\otimes N}\otimes E)$ and
the dot in the RHS stands for the inner products in the fibers of $S$ and $E$.

Next, one introduces the Dirac operator $D^{(E)}$ that acts on 
$\Gamma(S\otimes L^{\otimes N}\otimes E)$. Its action on 
    $\psi \in \Gamma(S\otimes L^{\otimes N}\otimes E)$ is defined by
    \begin{align}
      D^{(E)}\psi &= i \gamma^a \nabla_a \psi \notag\\
      &=i\gamma^a e^\mu_a \left(\partial_{\mu} +\frac{1}{4} \Omega_{\mu,bc} \gamma^{bc}-iNA_\mu -iA^{(E)}_\mu \right)\psi \ , \label{eq:dirac_op} 
    \end{align}
where $\gamma^a \; (a=1,2)$ are the gamma matrices satisfying
$\qty{ \gamma^a,\gamma^b} = 2 \delta^{ab}$ and $\gamma^{ab}= \frac{1}{2}[\gamma^a,\gamma^b]$.
$\Omega_{\mu,bc}$ are the spin connection and $A^{(E)}$ is the 
connection of $E$. In what follows, for concreteness, we put 
$\gamma^1=\sigma^1$
and $\gamma^2=\sigma^2$.
Then, the Dirac operator is decomposed on $S$ as follows:
    \begin{align} \label{eq:dirac_op_decompose}
      D^{(E)} = 
      \begin{pmatrix}
      0 & D^{(E)}_-\\
      D^{(E)}_+ & 0  
      \end{pmatrix}
    \end{align}
with $D^{(E)}_{\pm}=\nabla_1\pm i\nabla_2$.
The zero modes of the Dirac operator play 
a crucial role in the BT quantization.
It is known that $D^{(E)}_-$ does not have any normalizable zero modes
for sufficiently large $N$.
Thus, normalizable zero modes of 
$D^{(E)}$ are given by those of $D^{(E)}_+$, and the index theorem
tells that the number of the independent ones is $d^{(E)}N+c^{(E)}$,
where $d^{(E)}$ and $c^{(E)}$ are the rank and
the 1st Chern number of $E$, respectively.
We denote the set of all normalized zero modes of $D^{(E)}$ by 
$\mathrm{Ker} D^{(E)}$.

Let $\Pi$ and $\Pi'$ be projections 
on $\mathrm{Ker}D^{(E)}$ and $\mathrm{Ker}D^{(E')}$, respectively.
The Toeplitz operator $T^{(E',E)}(\chi)$ of $\chi \in \Gamma(\mathrm{Hom}(E,E'))$
is defined by projecting $\chi$ on the kernels of the Dirac 
operators as
\begin{equation}
    T^{(E',E)}(\chi) = \Pi' \chi \Pi \ .
\end{equation}
The BT quantization is the procedure for obtaining the Toeplitz operator.  
Suppose that $\qty{\psi_I|I = 1, \dots ,d^{(E)}N+c^{(E)}}$ and 
$\qty{\psi'_{J}|J = 1, \dots ,d^{(E')}N+c^{(E')}}$ are orthonormal 
bases of $\mathrm{Ker}D^{(E)}$ and $\mathrm{Ker}D^{(E')}$,
respectively.
Then, by using \eqref{eq:inner_product}, the Toeplitz operator
can be represented as 
a $(d^{(E')}N + c^{(E')})\times (d^{(E)}N + c^{(E)})$ matrix whose $(J,I)$ component is given by
    \begin{align}
      T ^{(E',E)}_{JI}(\chi) = (\psi'_{J},\chi \psi_I) \ . \label{eq:def_toeplitz_matrix}
    \end{align}

Finally, we state the properties of the Toeplitz operator in
the $N\to\infty$ limit.
The product of $T^{(E',E)}(\chi) = \Pi ' \chi \Pi$ and $T^{(E'',E')}(\chi') = \Pi'' \chi' \Pi'$
has the following asymptotic expansion in terms of $\hbar_N=V/N$
as $N\to\infty$ \cite{Adachi:2021ljw}:
    \begin{equation}\label{eq:aymptotic_expansion_Toeplitz}
      T^{(E'',E')}(\chi')T^{(E',E)}(\chi) = \sum_{i=0}^{\infty} \hbar_N ^ i T^{(E'',E)} (C_i(\chi',\chi)) \ , 
    \end{equation}
where $C_i$ are bi-linear differential operators with the order of derivative 
being at most $i$ that induce maps from
$(\mathrm{Hom}(E',E''),\mathrm{Hom}(E,E'))$ to $\mathrm{Hom}(E,E'')$.
The concrete forms of $C_i$ for $i=0,1,2$ are given by
    \begin{align}
      C_0 (\chi',\chi) =& \chi'\chi  \ , \label{eq:c0}\\
      C_1 (\chi',\chi) = &-\frac{1}{2}(g^{ab} + iW^{ab}) (\nabla_a \chi')(\nabla_b \chi)  \ , \label{eq:c1}\\
      C_2 (\chi',\chi) = &\frac{1}{8}(g^{ab} + iW^{ab}) (\nabla_a \chi')(R+4F^{(E')}_{12})(\nabla_b \chi) \notag \\
      &+\frac{1}{8}(g^{ab} + iW^{ab})(g^{cd} + iW^{cd})(\nabla_a \nabla_c \chi')(\nabla_b \nabla_d \chi) \label{eq:c2}\ ,
    \end{align}
where $W_{ab}=\epsilon^{ab}/\sqrt{g}$ are the Poisson tensor, $R$ is 
the scalar curvature and $F^{(E')}_{12} = e^\mu_1 e^\nu_2 F_{\mu\nu}=e^\mu_1 e^\nu_2(\partial_\mu A^{(E')}_\nu - \partial_\nu A^{(E')}_\mu)$ are the field strength of $E'$ in the
local Lorentz frame.
It follows from this asymptotic expansion that for
$\chi\in \mathrm{Hom}(E,E')$, $\chi'\in \mathrm{Hom}(E',E'')$ 
and a function $f$ on $M$
    \begin{align}
      &\lim_{N\to\infty} \abs{T(\chi')T(\chi) - T(\chi'\chi) } = 0 \label{eq:tp_product_N_to_infinity} \ ,\\
      &\lim_{N\to\infty} \abs{\hbar_N^{-1} [T(f\mathbbm{1}),T(\chi)]^{(E',E)}+iT^{(E',E)}(\qty{f,\chi})} = 0  \, \label{eq:tp_commutator_N_to_infinity}
      \ ,
    \end{align}
where a generalized commutator is defined by
    \begin{equation}\label{eq:def_commutator}
      [T(f\mathbbm{1}),T(\chi)]^{(E',E)} = T^{(E',E')}(f\mathbbm{1}_{E'})T^{(E',E)}(\chi) - T^{(E',E)}(\chi)T^{(E,E)}(f\mathbbm{1}_{E}) 
    \end{equation}
with $\mathbbm{1}_E$ being the identity
operator on $E$ and a generalized Poisson bracket is defined by
    \begin{equation}\label{eq:def_poisson_bracket}
      \qty{f,\chi} = W^{ab} (\partial_a f )(\nabla_b \chi)\ .
    \end{equation}

\section{Regularization of covariant derivatives as finite-size matrices}
\label{sec:matrix_regularization_for_covariant_derivatives}
As reviewed in section \ref{sec:covariant_derivative_interpretation},
in the covariant derivative interpretation,
$A_{(a)}$ are matrices with infinite size, since
$\nabla_{(a)}$ are  first-order differential operators
acting on $\Gamma(E_{reg})$.
Regularization is needed to remove divergences that
occur in the calculations of physical quantities \cite{Hanada:2006ei}. 
In particular, it is necessary to make the size of matrices finite
in order to apply the interpretation to nonperturbative calculations 
such as numerical simulations.

In this section, we study regularization of matrices
in the covariant derivative interpretation of matrix models.
By using the BT quantization, 
we represent the covariant derivatives in terms of finite-size matrices. 
We have seen in section \ref{subsec:BT_quantization} that the commutator of
the Toeplitz operators corresponds to the generalized Poisson 
bracket. 
In section \ref{subsec:matrix_regularization_for_covariant_derivatives}, by noting that 
the generalized Poisson bracket includes the covariant 
derivative, we construct matrices corresponding to
the covariant derivatives.
We restrict ourselves to the case of closed Riemann surfaces.
In section \ref{subsec:matrix_regularization_S2} and \ref{subsec:matrix_regularization_T2},
as examples, we consider $S^2$ and $T^2$ again and verify that the results in section \ref{subsec:hkk_S2} are reproduced.

\subsection{Representing covariant derivatives as finite-size matrices 
\label{subsec:matrix_regularization_for_covariant_derivatives}}
First, we explain how we treat tensor fields on a closed
Riemann surface. The behavior of 
the tensor fields in the local Lorentz frame can be 
classified by the U(1) charge.
For instance, the covariant derivative acts on
a spinor field $\psi={}^T(\psi^+,\psi^-)$ as
    \begin{align}
      \nabla_a \psi = 
        \begin{pmatrix}
          e^\mu_a(\partial_\mu - i (-\frac{1}{2}) \Omega_\mu{}^{12})\psi^+ \\
          e^\mu_a(\partial_\mu - i (\frac{1}{2}) \Omega_\mu{}^{12})\psi^-
        \end{pmatrix}
        \label{covariant derivative acting on spinor}\ .
    \end{align}
Thus, $\psi^{\pm}$ can be regarded as a field
with the U(1) charge, $s=\mp \frac{1}{2}$, respectively.
Moreover, we define $V^{\pm} = V^1 \pm iV^2$ for a vector field $V^a$. Then, the covariant derivative acts on $V^\pm$ as
    \begin{equation}
      \nabla_a V^\pm = e^\mu_a (\partial_\mu - i(\pm1)\Omega_\mu{}^{12})V^\pm \ .
    \end{equation}
The components $V^\pm$ of a contravariant vector field therefore
possess the charge, $s=\pm 1$, respectively.
It is also easy to see that the components of the covariant vector field $V_\pm$ possess the charge, $s=\mp 1$, respectively.
The components of the metric $\delta_{ab}$ are $\delta_{++}=\delta_{--}=0$ and $\delta_{+-}=\delta_{-+}=1/2$.
Hence, $V_{\pm}$ are given by $V_{\pm}=\delta_{\pm a}V^a=\frac{1}{2}V^{\mp}$. 

Let $E$ and $E'$ be fiber bundles with the connections being
$s\Omega_\mu{}^{12}$ and $s'\Omega_\mu{}^{12}$, respectively.
It follows from the above observation that
a component of a tensor field, $\chi$, 
that possesses the charge $q=s'-s$
can be interpreted as an element of $\mathrm{Hom}(E,E')$.
In this interpretation, the action of $\chi$ on $\psi \in \Gamma(E)$
is given by the simple product.
For a given $q$,
we can consider arbitrary $s$ and $s'$ such that
$s'-s=q$.

For Spin(2), irreducible representations are labeled by
the spin $s$ as in \eqref{spin s representation}, and 
the irreducible decomposition \eqref{Peter-Weyl} of the 
regular representation 
is given by the Fourier expansion 
where the dimension of each irreducible representation is one.
Thus, an element $\varphi(x,\theta)$ of $\Gamma(E_{reg})$ 
is expanded on a local patch as 
\begin{align}
\varphi(x,\theta) = \sum_q \varphi^{(q)}(x) e^{2iq\theta} \ ,
\label{Fourier expansion}
\end{align}
where $\theta$ is the coordinate of Spin(2) defined 
in \eqref{spin s representation} and the Fourier mode
$\varphi^{(q)}$ has the charge $q$.

\begin{figure}[t]
    \centering
    \includegraphics[width=7cm]{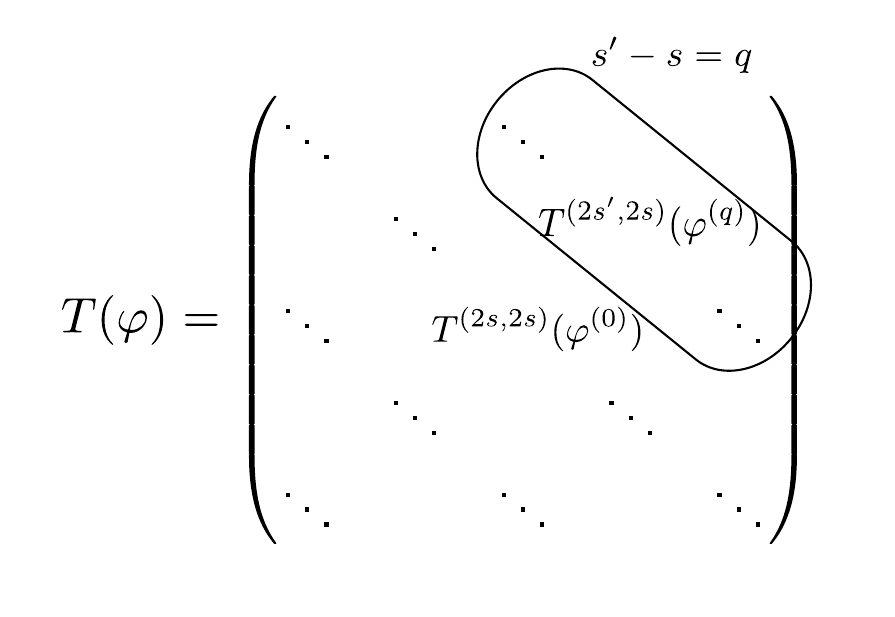}
    \caption{The operator $T(\varphi)$ whose $(2s',2s)$ block is 
    the Toeplitz operator of the spin $q=s'-s$ Fourier 
    component of an element
    $\varphi$ of $\Gamma(E_{reg})$.
    }
    \label{fig:1}
\end{figure}

In what follows, we replace the label of the fibers $(E',E)$
in the Toeplitz operator with $(2s',2s)$.
For a given element $\varphi$ of $\Gamma(E_{reg})$,
we define a matrix whose 
$(2s',2s)$ block is given by $T^{(2s',2s)}(\varphi^{(q)})$ with
$q=s'-s$ where $\varphi^{(q)}$ is the Fourier mode defined 
in \eqref{Fourier expansion}
and denote it by $T(\varphi)$ (Fig. \hspace{-0.1cm}\ref{fig:1})\footnote{
We can also define $T(\varphi)$  from the following viewpoint.
We begin with $\Gamma(E_{reg}\otimes E_{reg})$.
An element $\tilde{\varphi}(x,\theta',\theta)$ of 
$\Gamma(E_{reg}\otimes E_{reg})$ is expanded on a
local patch as $\tilde{\varphi}
(x,\theta',\theta)=\sum_{s',s}\varphi^{(s',s)}
(x)e^{-2is'\theta'+2is\theta}$.
Then, we can define the Toeplitz operators $T^{(2s',2s)}
(\varphi^{(s',s)})$.
From
\eqref{isomorphism}, we obtain
$\Gamma(E_{reg}\otimes E_{reg})=\Gamma(E_{reg})\oplus \Gamma(E_{reg})
\oplus\cdots$. 
This isomorphism is constructed using 
\eqref{isomorphism for regular rep} as
$\varphi(x,\theta',\theta)=\tilde{\varphi}(x,\theta'-\theta,\theta)$.
Thus, $s'$ labels an infinite copies of $\Gamma(E_{reg})$
in the RHS of the above equation. By putting $\varphi^{(s',s)}=\varphi^{(s)}$, we obtain $T(\varphi)$.
}.
If the range of $s$ and $s'$ is not restricted,
the size of $T(\varphi)$ is infinite, while that of each block
$T^{(2s',2s)}(\varphi^{(s'-s)})$ is finite.
We introduce a cut-off $\Xi$ for $s$ and $s'$
as $|s|, |s'| \leq \Xi$ in order to make the matrix size finite.
Eventually, we take the limit in which $N\rightarrow\infty$ and 
$\Xi\rightarrow\infty$ with $\Xi \ll N$ kept.
The blocks $T^{(2s',2s)}(\varphi^{(s'-s)})$ located on a line
parallel to the diagonal line possess the
same charge $q=s'-s$.
In this way, $T(\varphi)$ can be interpreted as a regularization
of the element $\varphi$ of $\Gamma(E_{reg})$.

Then, \eqref{eq:def_commutator} can be viewed as
the $(2s',2s)$ block of the commutator between $T(f\mathbbm{1})$
and $T(\varphi)$, where the $(2s,2s)$ block of $T(f\mathbbm{1})$
is given by $T^{(E,E)}(f\mathbbm{1}_{E})$ and its other off-diagonal
blocks vanish.

Let us consider a regularization of the
covariant derivatives. Here, we use 
an embedding of $M$ into $\mathbb{R}^n$ $(n\geq 3)$.
Let $X^A(A=1,\dots,n)$ be embedding coordinates.
The induced metric
$\delta_{ab}$ on $M$ in the Lorentz frame is given by 
$\delta_{ab}=\partial_aX^A\partial_bX^A$ 
where $\partial_a = e^\mu_a \partial_\mu$.
We see from \eqref{eq:tp_commutator_N_to_infinity}
that the generalized commutator \eqref{eq:def_commutator} of the Toeplitz operators corresponds to the generalized Poisson bracket \eqref{eq:def_poisson_bracket} on $M$.
Thus, in order to regularize the covariant derivatives as finite-size matrices, we define linear operators $\mathcal{P}_a$ that act on
the Toeplitz operator $T(\varphi)$ as follows:
    \begin{equation}\label{eq:def_Ppm}
      \mathcal{P}_\pm T(\varphi) = \pm i \hbar^{-1}_N T(\partial_\pm X^A) [T(X^A),T(\varphi)] \mp \frac{i}{2}\hbar^{-1}_N [T(\partial_\pm X^A),T(X^A)]T(\varphi) \ ,
    \end{equation}
where $\hbar_N=V/N$.

\begin{figure}[t]
    \centering
    \includegraphics[width=15cm]{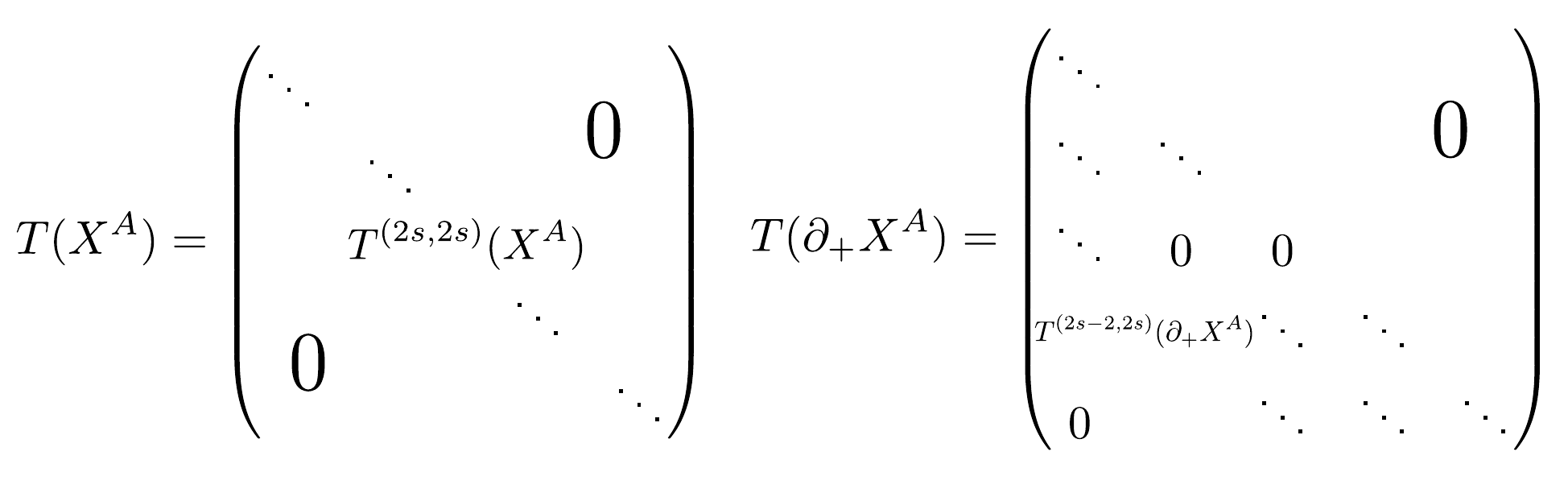}
    \caption{The structure of $T(X^A)$ and $T(\partial_+ X^A)$.}
    \label{fig:2}
\end{figure}

\eqref{eq:def_Ppm} is rewritten in terms of block components as
    \begin{equation}
      \big(\mathcal{P}_\pm T(\varphi)\big)^{(2s',2s)}_{mn} = \sum_{t,u}\sum_{k,l} \big(\mathcal{P}_\pm\big)^{(2s',2s)(t,u)}_{(mn)(kl)} \big(T(\varphi)\big)^{(t,u)}_{kl} \ .
      \label{P_pm block}
    \end{equation}
Here $(2s',2s)$ and $(m,n)$ correspond to the row index, while
$(t,u)$ and $(k,l)$ to the column index.
In this sense, $\mathcal{P}_{\pm}$ are matrices with finite size 
acting on $T(\varphi)$.
The non-zero blocks in $T(X^A)$ and $T(\partial_\pm X^A)$ are 
$(2s,2s)$ and $(2s\mp2,2s)$, respectively 
(Fig.\hspace{-0.1cm} \ref{fig:2}).
In \eqref{P_pm block}, therefore, the only block components with 
$2s'-t=\mp 2$ and $2s=u$ of $\mathcal{P}_{\pm}$ are
non-vanishing.

We examine the large-$N$ behavior of $\mathcal{P}_\pm T(\varphi)$ 
defined in \eqref{eq:def_Ppm}
by using the asymptotic properties of the Toeplitz 
operator, \eqref{eq:tp_product_N_to_infinity}
and \eqref{eq:tp_commutator_N_to_infinity}.
The first term in the RHS of \eqref{eq:def_Ppm}
is evaluated as follows:
\begin{align}
      \pm i \hbar_N^{-1}T(\partial_\pm X^A )[T(X^A),T(\varphi)] =& T(\pm \partial_{\pm} X^A \epsilon^{ab}\partial_a X^A \nabla_b \varphi) + \mathcal{O}\qty(\frac{1}{N})\notag\\ 
      =&T(\pm \delta_{\pm a}\epsilon^{ab}\nabla_b\varphi)+ \mathcal{O}\qty(\frac{1}{N})\notag\\
      =& T(i\nabla_\pm \varphi) + \mathcal{O}\qty(\frac{1}{N}) \ . \label{eq:Ppm_cov_der}
    \end{align}
This is what we expected.
On the other hand,
the second term in the RHS of 
\eqref{eq:def_Ppm} is evaluated as
    \begin{align}
      i \hbar_N^{-1} [T(\partial_\pm X^A), T(X^A)]T(\varphi) =& T(\qty{\partial_\pm X^A,X^A}\varphi) + \mathcal{O}\qty(\frac{1}{N}) 
      \ .
    \end{align}
Here we further calculate $\qty{\partial_\pm X^A,X^A}$:
    \begin{align}
      \{ \partial_+ X^A , X^A\} 
      =& -2i((\nabla_+ \partial_+ X^A) \partial_- X^A - (\nabla_- \partial_+ X^A)\partial_+ X^A )\notag\\
      =& -2i(\nabla_+(\partial_+X^A \partial_-X^A)-\partial_+ X^A \nabla_+\partial_-X^A -\frac{1}{2}\nabla_-(\partial_+X^A \partial_+ X^A)) \notag \\
      =& 2i\partial_+ X^A \nabla_-\partial_+ X^A \notag\\
      =&0 \ .
    \end{align}
In the third equality, we used the facts that the induced metric
in the local Lorentz frame is given by $\delta_{ab} = \partial_a X^A \partial_b X^A$ and that 
$\nabla_+ \partial_- X^A = \nabla_- \partial_+ X^A$.
In the same way, we can show that
$\{ \partial_- X^A , X^A\}=0$.
Thus, 
the second term in the RHS of \eqref{eq:def_Ppm}
vanishes as $\mathcal{O}(N^{-1})$ in the large-$N$ limit.

In fact, we introduce this term in order that
$(\mathcal{P}_\pm)^\dagger = \mathcal{P}_\mp$ holds for finite $N$.
By using the Frobenius inner product, 
we define the inner product between two Toeplitz operators
$T(\varphi)$ and $T(\varphi')$ as
    \begin{align}
      (T(\varphi),T(\varphi ')) =& \mathrm{Tr}((T(\varphi))^\dagger 
      T(\varphi '))\notag \\
      =& \sum_{s',s} \sum_{m,n} \Big(\big(T(\varphi)\big)^{(2s',2s)}_{mn}\Big)^*
      \big(T(\varphi ')\big)^{(2s',2s)}_{mn} \ .
    \end{align}
By using this inner product, we define the Hermite conjugate 
$(\mathcal{P}_\pm)^\dagger$ of $\mathcal{P}_\pm$  by
$((\mathcal{P}_\pm)^\dagger T(\varphi),T(\varphi '))=(T(\varphi),\mathcal{P}_\pm T(\varphi '))$.
Then, we see that 
\begin{align}
      &((\mathcal{P}_{+})^\dagger T(\varphi),T(\varphi')) \notag\\
      &= (T(\varphi),\mathcal{P}_{+}T(\varphi')) \notag \\
      &= i\hbar_N^{-1}\mathrm{Tr}\Big( (T(\varphi))^\dagger T(\partial_{+}X^A)[T(X^A),T(\varphi')]- \frac{1}{2}(T(\varphi))^\dagger[T(\partial_{+}X^A),T(X^A)]T(\varphi')\Big) \notag \\
      &= i\hbar_N^{-1}\mathrm{Tr}\Big((T(\varphi))^\dagger T(\partial_{+}X^A)T(X^A)T(\varphi')-  (T(\varphi))^\dagger T(\partial_{+}X^A)T(\varphi')T(X^A)\notag\\
      &\,\,\,\,\,\,\,- \frac{1}{2}(T(\varphi))^\dagger[T(\partial_{+}X^A),T(X^A)]T(\varphi')\Big) \notag \\
      &=i\hbar_N^{-1}\mathrm{Tr}\Big( (T(\varphi))^\dagger T(X^A)T(\partial_{+}X^A)T(\varphi')- T(X^A)(T(\varphi))^\dagger T(\partial_{+}X^A)T(\varphi')\notag\\
      &\,\,\,\,\,\,\,+ \frac{1}{2}(T(\varphi))^\dagger[T(\partial_{+}X^A),T(X^A)]T(\varphi')\Big) \notag \\
      &=i\hbar_N^{-1}  \mathrm{Tr}\Big(  [(T(\varphi))^\dagger, T(X^A)]T(\partial_{+}X^A)T(\varphi')+ \frac{1}{2}(T(\varphi))^\dagger[T(\partial_{+}X^A),T(X^A)]T(\varphi')\Big) \notag \\
      &=\mathrm{Tr}\bigg( \Big( - i \hbar_N^{-1}T(\partial_{-}X^A)[T(X^A),T(\varphi)]\Big)^{\dagger}T(\varphi')+\Big(\frac{i}{2}\hbar_N^{-1}[T(\partial_{-}X^A),T(X^A)]T(\varphi) \Big)^\dagger T(\varphi')\bigg)\notag \\
      &=(\mathcal{P}_{-}T(\varphi),T(\varphi')) \ .
\end{align} 
In the third equality, we expanded the commutator in the first term.
In the forth equality, we exchanged $T(\partial_+ X^A)$ and $T(X^A)$ in 
the first term, changed the sign of the third term 
and used the cyclic property of the trace in the second term.
In the fifth equality, we put the first and second terms 
together in the commutator.
Thus, we see that 
$(\mathcal{P}_+)^\dagger=\mathcal{P}_-$ for finite $N$.
Similarly, we see that $(\mathcal{P}_-)^\dagger=\mathcal{P}_+$.

In this way, $\mathcal{P}_\pm$ defined in \eqref{eq:def_Ppm} 
are finite-size matrices that serves as 
a regularization of the covariant derivatives.
Note that $\mathcal{P}_{\pm}$ is local Lorentz invariant.
The Toeplitz operators are defined as \eqref{eq:def_toeplitz_matrix} 
with a inner product that is local Lorentz invariant.
Then, the U(1) charges of $X^A$ and $\partial_\pm X^A$ are conserved.
Thus, the index $\pm$ of $\mathcal{P}_\pm$ is not transformed under 
the local Lorentz transformation.
This implies that the index of $\mathcal{P}_{\pm}$ is equivalent 
to that of the covariant derivative $\nabla_{(a)}$ acting 
on $\Gamma(E_{reg})$.
To summarize, $\mathcal{P}_a$ are finite-size matrices corresponding to $i\nabla_{(a)}$.

By using the commutator between $\mathcal{P}_+$ and $\mathcal{P}_-$,
we also define $\mathcal{P}_3$:
    \begin{align}
      \mathcal{P}_3 T(\varphi) =& [\mathcal{P}_+ , \mathcal{P}_-]T(\varphi) \notag\\
      =&\mathcal{P}_+ (\mathcal{P}_-T(\varphi)) - \mathcal{P}_- (\mathcal{P}_+ T(\varphi))\ .\label{eq:def_P3}
    \end{align}
Because the second term of \eqref{eq:def_Ppm} is $\mathcal{O}(N^{-1})$, 
we ignore this term below.
The first term in the RHS of \eqref{eq:def_P3} is evaluated as
\begin{align}
    \mathcal{P}_+ (\mathcal{P}_-T(\varphi)) 
    =& - \mathcal{P}_+ (i \hbar_N^{-1} T(\partial_- X^A)[T(X^A),T(\varphi)]) + \mathcal{O}\qty(\frac{1}{N}) \notag \\
    =& \hbar_N^{-2} T(\partial_+ X^B)[T(X^B),T(\partial_-X^A)[T(X^A),T(\varphi)]] + \mathcal{O}\qty(\frac{1}{N})\notag \\
    =& \hbar_N^{-2}T(\partial_+X^B)T(\partial_-X^A)[T(X^B),[T(X^A),T(\varphi)]] \notag\\
    &+\hbar_N^{-2}T(\partial_+X^B)[T(X^B),T(\partial_-X^A)][T(X^A),T(\varphi)] + \mathcal{O}\qty(\frac{1}{N})\ .
\end{align}
Similarly, the second term in the RHS of  
\eqref{eq:def_P3} is evaluated.
Then, $\mathcal{P}_3$ is evaluated as
\begin{align}
    \mathcal{P}_3 T(\varphi) 
    =& \hbar_N^{-2}\Big( T(\partial_+ X^B) T(\partial_- X^A) [T(X^B),[T(X^A),T(\varphi)]] \notag\\
    &+ T(\partial_+ X^B)[T(X^B),T(\partial_-X^A)][T(X^A),T(\varphi)]\notag \\
    &-T(\partial_- X^A) T(\partial_+ X^B) [T(X^A),[T(X^B),T(\varphi)]] \notag\\
    &- T(\partial_- X^A)[T(X^A),T(\partial_+X^B)][T(X^B),T(\varphi)]\Big)+\mathcal{O}\qty(\frac{1}{N})\notag \\
    =& \hbar_N^{-2}\Big( T(\partial_+ X^B) T(\partial_- X^A) [T(X^B),[T(X^A),T(\varphi)]]\notag\\
    &+ T(\partial_+ X^B)[T(X^B),T(\partial_-X^A)][T(X^A),T(\varphi)]\notag \\
    &-T(\partial_+ X^B) T(\partial_- X^A) [T(X^A),[T(X^B),T(\varphi)]] \notag\\
    &- T(\partial_- X^A)[T(X^A),T(\partial_+X^B)][T(X^B),T(\varphi)]\Big)+\mathcal{O}\qty(\frac{1}{N})\notag \\
    =& \hbar_N^{-2} \Big( T(\partial_+ X^B) T(\partial_- X^A) [[T(X^A),T(X^B)], T(\varphi)]\notag\\
    &+ T(\partial_+ X^B)[T(X^B),T(\partial_-X^A)][T(X^A),T(\varphi)]\notag \\
    &- T(\partial_- X^A)[T(X^A),T(\partial_+X^B)][T(X^B),T(\varphi)]\Big)+\mathcal{O}\qty(\frac{1}{N})\ . \label{eq:P3_largeN_Jacobi}
\end{align}
In the second equality, using \eqref{eq:tp_product_N_to_infinity}, we exchanged $T(\partial_-X^A)$ and $T(\partial_+X^B)$ in the third term. 
In the third equality, we used the Jacobi identity for the first and third terms.
Furthermore, using \eqref{eq:tp_product_N_to_infinity}, \eqref{eq:tp_commutator_N_to_infinity} 
and \eqref{eq:Ppm_cov_der}, we can evaluate
the second term in the most right-hand side of 
\eqref{eq:P3_largeN_Jacobi} as
\begin{align}
    &\hbar_N^{-2}T(\partial_+ X^B)[T(X^B),T(\partial_-X^A)][T(X^A),T(\varphi)] \notag\\
    &= \hbar_N^{-1} T(\nabla_+\partial_-X^A)T(\{X^A,\varphi\}) + \mathcal{O}\qty( \frac{1}{N} ) \notag \\
    &= \hbar_N^{-1} T(\nabla_+\partial_-X^A\{X^A,\varphi\}) + \mathcal{O}\qty( \frac{1}{N} )\ .
\end{align}
Finally, we calculate
$\nabla_+\partial_- X^A \{X^A,\varphi\}$ as follows:
    \begin{align}
      \nabla_+\partial_-X^A\{X^A,\varphi\} =& -2i\nabla_+\partial_-
      X^A(\partial_+X^A\nabla_-\varphi-\partial_-
      X^A\nabla_+\varphi)\notag\\  
      =&-2i\{ (\nabla_-\partial_+X^A)\partial_+X^A \nabla_-\varphi-
      (\nabla_+\partial_-X^A)\partial_-X^A \nabla_+\varphi\}\notag\\
      =&-i\{\nabla_-(\delta_{++})\nabla_-\varphi - \nabla_
      +(\delta_{--})\nabla_+\varphi\} \notag\\
      =&0\ .
    \end{align}
Hence, the second term in the most right-hand side of \eqref{eq:P3_largeN_Jacobi} behaves as $\mathcal{O}(N^{-1})$. 
Similarly, we can also evaluate
the third term in the RHS of \eqref{eq:P3_largeN_Jacobi}.
Thus, $\mathcal{P}_3$ behaves as    
\begin{equation}\label{eq:P3_N_to_infinity}
      \mathcal{P}_3T(\varphi)= \hbar_N^{-2} T(\partial_+ 
      X^A)T(\partial_- X^B)[[T(X^A),T(X^B)],T(\varphi)] 
      +\mathcal{O}\qty(\frac{1}{N}) \ .
      \end{equation}

  \subsection{Example: $S^2$} \label{subsec:matrix_regularization_S2}
    As in section \ref{subsec:hkk_S2},
    we consider a unit sphere $S^2$ embedded  
    in $\mathbb{R}^3$ and apply to it the formulation developed
    in the previous subsection.
    In particular, we see that
    $\mathcal{P}_a$ defined in the previous subsection
    form the su(2) Lie algebra as seen in section \ref{subsec:hkk_S2}.

    Let $X^A$ be again embedding coordinates of $S^2$ and
    $(z,\bar{z})$ be the stereographic projection of $S^2$ from
    the north pole to the $X^1-X^2$ plane:
    \begin{equation}
      X^1 = \frac{z+\bar{z}}{1+|z|^2} \ ,\; X^2 = -i\frac{z-\bar{z}}{1+|z|^2} \ , \; X^3 = \frac{1-|z|^2}{1+|z|^2} \ .
    \end{equation}
    The induced metric, the zweibein, the inverse of the 
    zweibein and the spin connection are given by
    \begin{align}
      &g_{z\bar{z}} =g_{\bar{z}z} = \frac{2}{(1+|z|^2)^2}  \ , \; \; 
      g_{zz}=g_{\bar{z}\bar{z}}=0 \ , \notag\\
      &e^1 = \frac{dz + d\bar{z}}{1+|z|^2}\ ,\;\; e^2 = -i \frac{dz - d\bar{z}}{1+|z|^2}  \ , \notag\\
      &e_1 = \frac{1}{2} (1+|z|^2) ( \partial_z + \partial_{\bar{z}}) \
      , \; \; e_2 = \frac{i}{2} (1+|z|^2) ( \partial_z - 
      \partial_{\bar{z}}) \ , \notag\\
      &\Omega^1{}_2 = -i \frac{\bar{z}dz - z d\bar{z}}{1+|z|^2} \ ,
    \end{align}
    as in section \ref{subsec:hkk_S2}. The volume form is given by
    \begin{align}
      \omega =& i \sqrt{\det g} dz \wedge d\bar{z} \notag \\
      =& i \frac{2}{(1+|z|^2)^2} dz \wedge d\bar{z} \ ,
    \end{align}
    and the volume is given by
    \begin{equation}
      V = \frac{1}{2\pi} \int_{S^2} \omega = 2 \ .
    \end{equation}
    Next, as a connection 1-form in the complex line bundle $L$,
    we adopt the Wu-Yang monopole, 
    \begin{equation}
      A = - \frac{i}{2} \frac{\bar{z}dz - z d\bar{z}}{1+|z|^2} \ ,
    \end{equation}
    which satisfies that $F=dA =\omega/V$. Note that 
    $A=\frac{1}{2}\Omega^1{}_2 $.
    The connections in the fiber bundles $E$ and $E'$ are given by
    \begin{align}
      A^{(E)} =& s \Omega^1{}_2 \ , \notag\\
      A^{(E')} =& s' \Omega^1{}_2 \ ,
    \end{align}
    respectively.
    The Dirac operator $D^{(E)}$ on 
    $\Gamma(S\otimes L^{\otimes N}\otimes E)$ reads
    \begin{align} \label{eq:dirac_op_S2}
      D^{(E)} =& 
      \begin{pmatrix}
      0 & D^{(E)}_-\\
      D^{(E)}_+ & 0  
      \end{pmatrix}
    \end{align}
with
     \begin{align}
      D_-^{(E)} =& i \qty{(1+|z|^2)\partial_z - \frac{N+2s+1}{2}\bar{z}} \ , \notag\\
      D_+^{(E)} =& i \qty{(1+|z|^2)\partial_{\bar{z}} + \frac{N+2s-1}{2}z} \ .
    \end{align}
    For $N \geq 2s+1$, $D^{(E)}_-$ has no normalizable
    zero modes so that $\mathrm{Ker}D^{(E)}_-$ is empty.
    In what follows, we assume that this condition is satisfied. 
    Then, the orthonormalized zero modes
    take the following form:
    \begin{align}
      &\psi^{(E)}_n = \begin{pmatrix} \psi^{(E)+}_n \\0 \end{pmatrix} ,\\
      &\psi^{(E)+}_n = \sqrt{\frac{N+2s}{4\pi}}\binom{N+2s-1}{n}^\frac{1}{2} \frac{z^n}{(1+|z|^2)^{\frac{N+2s-1}{2}}} \ ,
    \end{align}
    where $n=0,1,\dots,N+2s-1$. 
    By using the formula,
    \begin{equation}
      \frac{1}{2\pi}\int idzd\bar{z} \frac{z^a\bar{z}^b}{(1+|z|^2)^c} = \delta_{ab}\frac{\Gamma(a+1)\Gamma(c-a-1)}{\Gamma(c)} \ , 
    \end{equation}
    where $\Gamma(a)$ is the Gamma function, we can calculate the 
    Toeplitz operators of $X^A$.
    The result is 
    \begin{align}
      \big(T(X^+)\big)^{(2s,2s)}_{mn} &= \delta_{m-1,n} \frac{2}{N+2s+1}\sqrt{m(N+2s-m)} \ ,\notag\\
      \big(T(X^-)\big)^{(2s,2s)}_{mn} &=\Big(\big(T(X^+)\big)^{(s,s)}_{mn}\Big)^\dagger \notag\\
      &= \delta_{m+1,n} \frac{2}{N+2s+1}\sqrt{(m+1)(N+2s-m-1)} 
      \ ,\notag\\
      \big(T(X^3)\big)^{(2s,2s)}_{mn} &= \delta_{m,n} \frac{N+2s-2n-1}{N+2s+1} \ ,
    \label{Toeplitz operator of X^A}
    \end{align}
    where $m,\ n=0,\ldots , N+2s-1$.
    These operators satisfy the su(2) Lie algebra:
    \begin{equation} \label{eq:toeplitz_su(2)}
      [T(X^A),T(X^B)]^{(2s,2s)} = - \frac{2}{N+2s+1} i \epsilon^{ABC} \big(T(X^C)\big)^{(2s,2s)} \ .
    \end{equation}
   Furthermore, noting that $\partial_\pm X^A$ possess the charge 
   $q=\mp 1$, we can calculate the Toeplitz operators of 
   $\partial_\pm X^A$ as
    \begin{align}
      &\big(T(\partial_+ X^+)\big)^{(2s-2,2s)}_{m'n} = 
      \delta_{m',n}\sqrt{\frac{(N+2s-n-1)(N+2s-n-2)}{(N+2s)(N+2s-1)}}
      \ ,\notag\\
      &\big(T(\partial_+ X^-)\big)^{(2s-2,2s)}_{m'n} = -
      \delta_{m'+2,n}\sqrt{\frac{n(n-1)}{(N+2s)(N+2s-1)}} \ ,\notag\\
      &\big(T(\partial_+ X^3)\big)^{(2s-2,2s)}_{m'n} = -
      \delta_{m'+1,n}\sqrt{\frac{n(N+2s-n-1)}{(N+2s)(N+2s-1)}} 
      \ ,\notag\\
      &\big(T(\partial_- X^+)\big)^{(2s+2,2s)}_{m''n} = -
      \delta_{m'',n+2}\sqrt{\frac{m''(m''-1)}{(N+2s+2)(N+2s+1)}}
      \ ,\notag\\
      &\big(T(\partial_- X^-)\big)^{(2s+2,2s)}_{m''n} = 
      \delta_{m'',n}\sqrt{\frac{(N+2s-m''+1)(N+2s-m'')}{(N+2s+2)
      (N+2s+1)}} \  ,\notag\\
      &\big(T(\partial_- X^3)\big)^{(2s+2,2s)}_{m''n} = -
      \delta_{m'',n+1}\sqrt{\frac{m''(N+2s-m''+1)}{(N+2s+2)(N+2s+1)}} 
      \ ,
    \label{Toeplitz operator of partial X^A}
    \end{align}
    where $m'=0,\ldots,N+2s-3$, $m''=0,\ldots,N+2s+1$ and $n=0,\ldots,N+2s-1$.
    
    By using \eqref{Toeplitz operator of X^A} and 
    \eqref{Toeplitz operator of partial X^A},
    we can calculate $\mathcal{P}_\pm$ and $\mathcal{P}_3$ that we defined in the previous subsection.
    In appendix \ref{app:P+-_S2}, we present concrete forms of $\mathcal{P}_\pm$.
    In appendix \ref{appA}, we evaluate $\mathcal{P}_3$ by using 
    \eqref{eq:P3_N_to_infinity}.
    It is given up to $\mathcal{O}(N^{-1})$ by
    \begin{align}
      \big(\mathcal{P}_3T(\varphi)\big)^{(2s',2s)}_{mn} = -\frac{s'-s}{2}\big(T(\varphi)\big)^{(2s',2s)}_{mn} \ .\label{eq:P3_on_S2}
    \end{align}
    As anticipated from \eqref{eq:nabla+_nabla-_commutator}
    and \eqref{eq:P3_N_to_infinity}, 
    $\mathcal{P}_3$ is proportional to the charge of fields. 
    Indeed, the charges are the eigenvalues of 
    the local Lorentz generator $O_{+-}$.

    We also see from this that the action of the commutators 
    between $\mathcal{P}_3$ and $\mathcal{P}_\pm$ on the Toeplitz operator 
    are given up to $\mathcal{O}(N^{-1})$ as 
    \begin{align}
      \big([\mathcal{P}_3,\mathcal{P}_+]T(\varphi)\big)^{(2s'-2,2s)} =& \frac{1}{2}\big(\mathcal{P}_+T(\varphi)\big)^{(2s'-2,2s)}  \ , \notag\\
      \big([\mathcal{P}_3,\mathcal{P}_-]T(\varphi)\big)^{(2s'+2,2s)} =& -\frac{1}{2}\big(\mathcal{P}_-T(\varphi)\big)^{(2s'+2,2s)} \ .
    \end{align}
    Thus, the commutators between $\mathcal{P}_\pm$ and $\mathcal{P}_3$ are summarized up to $\mathcal{O}(N^{-1})$ as
    \begin{align}
      &[\mathcal{P}_+,\mathcal{P}_-] = \mathcal{P}_3 \  , \\
      &[\mathcal{P}_3,\mathcal{P}_\pm] = \pm \frac{1}{2} \mathcal{P}_\pm \ ,
    \end{align}
     which are the su(2) Lie algebra.
     This reproduces 
   \eqref{eq:nabla+_nabla-_commutator} and \eqref{eq:lorentz_nabla_commutator}
   in section \ref{subsec:hkk_S2}. 

\subsection{Example: $T^2$} \label{subsec:matrix_regularization_T2}
In this subsection, we consider 
a torus $T^2$ embedded in $\mathbb{R}^4$.
See \cite{Adachi:2021aux} for the detailed derivation of the Dirac 
zero modes and the Toeplitz operators. 
As the orthonormal coordinates $(x^1,x^2)$ of $T^2$ satisfy 
the periodicity $x^a\sim x^a+2\pi$,
we introduce the fundamental functions $u$ and $v$ on $T^2$ as
\begin{align}
    u(x^1,x^2) = e^{ix^1},\quad v(x^1,x^2)=e^{ix^2}.
\end{align}
Then, the isometric embedding $X^A$ 
of $T^2$ in $\mathbb{R}^4$ is defined by
\begin{align}
    &X^1(x^1,x^2) = \cos{x^1} = \frac{1}{2}(u+u^*) \ , \; \; X^2(x^1,x^2)=\sin{x^1}=\frac{1}{2i}(u-u^*)\ ,\notag\\
    &X^3(x^1,x^2) = \cos{x^2} = \frac{1}{2}(v+v^*)\ , \; \; X^4(x^1,x^2)=\sin{x^2}=\frac{1}{2i}(v-v^*)\ .   
\end{align}
The derivatives of those satisfy
\begin{align}
    &\partial_1 X^1 = -X^2 \ , \; \; \partial_1 X^2 = X^1 \ ,\notag\\
    &\partial_2 X^3 = -X^4 \ , \; \; \partial_2 X^4 = X^3 \ ,\notag\\
    &\partial_2 X^1 = \partial_2 X^2 = \partial_1 X^3 = \partial_1 X^4 = 0\ .
\end{align}
The spin connection of $T^2$ is zero.
The volume form is given by
\begin{equation}
    \omega = dx^1 \wedge dx^2\ .
\end{equation}
The volume of $T^2$ is given by
\begin{equation}
    V=\frac{1}{2\pi} \int_{T^2} \omega = \frac{1}{2\pi}\int_{T^2} dx^1dx^2 = 2\pi\ ,
\end{equation}
and $\hbar_N =2\pi/N$.
We adopt the following U(1) connection $A$ of the line bundle $L$ as
\begin{equation}
    A=\frac{1}{4\pi}(-x^2dx^1 +x^1 dx^2)\ .
\end{equation}
This satisfies $F=dA= \omega / V$.
$A$ varies under the shift $x^a \to x^a+2\pi$ as
\begin{align}
    &A(x^1+2\pi,x^2) = A(x^1,x^2) + \frac{1}{2}dx^2 \ ,\notag\\
    &A(x^1,x^2+2\pi) = A(x^1,x^2) - \frac{1}{2}dx^1 \ .
\end{align}
This is considered as a gauge transformation.
Then, the spinor field $\psi \in \Gamma(S\otimes L^{\otimes N})$ satisfies the boundary condition,
\begin{align}
    &\psi(x^1+2\pi,x^2) = \exp\qty(i\frac{N}{2}x^2) \psi(x^1,x^2)\ ,\notag\\
    &\psi(x^1,x^2+2\pi) = \exp\qty(-i\frac{N}{2}x^1) \psi(x^1,x^2)\ .
\end{align}
The Dirac operator $D^{(E)}$ on 
$\Gamma(S\otimes L^{\otimes N}\otimes E)$ reads
    \begin{align} \label{eq:dirac_op_T2}
      D^{(E)} =& 
      \begin{pmatrix}
      0 & D^{(E)}_-\\
      D^{(E)}_+ & 0  
      \end{pmatrix}
    \end{align}
with
     \begin{align}
      D_-^{(E)} =& i (\partial_1 -i\partial_2 - \frac{N}{4\pi}(x^1-ix^2)) \ , \notag\\
      D_+^{(E)} =& i (\partial_1 +i\partial_2 + \frac{N}{4\pi}(x^1+ix^2)) \ .
    \end{align}
As in the case of $S^2$, 
for sufficiently large $N$, $D^{(E)}_-$ has no normalizable zero modes so that $\mathrm{Ker}D^{(E)}_-$ is empty.
Then, the orthonormalized zero modes of $D^{(E)}$ are given as follows:
\begin{align}
    &\psi^{(E)}_n = \begin{pmatrix} \psi^{(E)+}_n \\0 \end{pmatrix} ,\\
    &\psi^{(E)+}_n = d_ne^{-\frac{N}{4\pi}(x^2)^2}e^{i\frac{N}{4\pi}x^1x^2} \sum_{k\in \mathbb{Z}} e^{-\pi(Nk^2+2nk)}e^{i(Nk+n)(x^1+ix^2)} \ ,
\end{align}
where $n=1,\dots,N$. The normalization factor $d_n$ is given by
\begin{equation}
    d_n =\qty(\frac{N}{8\pi^4})^{1/4}e^{-\frac{\pi}{N}n^2}\ .        
\end{equation}
Note that the Dirac zero modes are independent of the charge of $E$.
When we perform the Berezin-Toeplitz quantization, we must be careful 
about the charge of the component of the tensor field.
For example, the embedding coordinate $X^1$ and the derivatives $2\partial_\pm X^2$ are the same function of $x^a$, but they have the charges $0$ and $\mp 1$, respectively.
Therefore, the non-zero blocks of the Toeplitz operators $T(X^1)$ and $T(\partial_\pm X^2)$ are $(2s,2s)$ and $(2s \mp 2 ,2s)$ blocks, respectively, and all these blocks are the same matrices.

Now, let us compute the Toplitz operators.
Suppose that $u$ and $v$ have the charge $s'-s$.
Then, the Toeplitz operators $\big(T(u)\big)^{(2s',2s)}$ and $\big(T(v)\big)^{(2s',2s)}$ 
are given by the clock-shift matrices $U$ and $V$ that take the 
following forms
\begin{align}
    U = e^{-\frac{\pi}{2N}}
    \begin{pmatrix}
       & &       & & 1 \\
      1& &       & &   \\
       &1&       & &   \\
       & &\ddots & &   \\
       & &       &1&   
    \end{pmatrix} \ , \; \;
    V = e^{-\frac{\pi}{2N}}
    \begin{pmatrix}
       q^{-1}&      &       & \\
             &q^{-2}&       & \\
             &      &\ddots & \\
             &      &       & q^{-N}    
    \end{pmatrix}  \label{eq:shift_clock}
\end{align}
with $q=e^{i2\pi/N}$.
These matrices satisfy the 't Hooft-Weyl algebra,
\begin{align}
    UV = qVU\ .  \label{eq:shift_clocl_algebra}
\end{align}
Hence, the non-zero components of 
the Toplitz operators of the embedding coordinates $X^A$ are given by
\begin{align}
    &\big(T(X^1)\big)^{(2s,2s)} = \frac{1}{2}(U+U^\dagger) \ , \; \; 
    \big(T(X^2)\big)^{(2s,2s)} = \frac{1}{2i}(U-U^\dagger) \ ,\notag \\
    &\big(T(X^3)\big)^{(2s,2s)} = \frac{1}{2}(V+V^\dagger) \ , \; \; 
    \big(T(X^4)\big)^{(2s,2s)} = \frac{1}{2i}(V-V^\dagger)\ ,
    \label{eq:Toeplitz_XA_T2}
\end{align}
and those of the derivatives $\partial_a X^A$ are given by
\begin{align}
    &\big(T(\partial_+ X^1)\big)^{(2s-2,2s)} = \big(T(\partial_- X^1)\big)^{(2s+2,2s)} = - \frac{1}{4i}(U-U^\dagger) \ ,\notag\\
    &\big(T(\partial_+ X^2)\big)^{(2s-2,2s)} = \big(T(\partial_- X^2)\big)^{(2s-2,2s)} = \frac{1}{4}(U+U^\dagger) \ ,\notag\\
    &\big(T(\partial_+ X^3)\big)^{(2s-2,2s)} = -\big(T(\partial_- X^3)\big)^{(2s+2,2s)} = \frac{1}{4}(V-V^\dagger) \ ,\notag\\
    &\big(T(\partial_+ X^4)\big)^{(2s-2,2s)} = -\big(T(\partial_- X^4)\big)^{(2s-2,2s)}= - \frac{i}{4}(V+V^\dagger) \ .
\end{align}
Substituting these into \eqref{eq:def_Ppm}, we obtain the non-zero blocks of $\mathcal{P}_\pm$ as
\begin{align}
    &(\mathcal{P}_+)^{(2s'-2,2s)(2s',2s)}_{(mn)(kl)} = -i\hbar_N^{-1} \big(T(\partial_+X^A)\big)^{(2s'-2,2s')}_{mk}\big(T(X^A)\big)^{(2s,2s)}_{ln} \ ,
    \notag \\
    &(\mathcal{P}_-)^{(2s'+2,2s)(2s',2s)}_{(mn)(kl)} = i\hbar_N^{-1} \big(T(\partial_-X^A)\big)^{(2s'+2,2s')}_{mk}\big(T(X^A)\big)^{(2s,2s)}_{ln} \ ,
\end{align}
where we used
\begin{equation}
    T(\partial_\pm X^A)T(X^A) = 0 \ .
\end{equation}
Note that the size of $\big(\mathcal{P}_\pm\big)^{(2s'\mp2,2s)(2s',2s)}$ on $T^2$ is independent of the charges $s$ and $s'$ unlike that on $S^2$.
Then, as shown in the appendix \ref{appA}, $\mathcal{P}_3T(\varphi)$ 
behaves as
\begin{equation}
    \mathcal{P}_3T(\varphi) = \mathcal{O}\qty(\frac{1}{N})
\end{equation} 
for sufficiently large $N$.
This reflects the flatness of $T^2$.

\section{Conclusion and discussion} \label{sec:conclusion}
In this paper, we studied regularization of matrices in 
the covariant derivatives interpretation of matrix models.
By using the Berezin-Toeplitz quantization,
we represented the covariant derivatives in two dimensions
in the covariant derivative interpretation 
as matrices with finite size.
We verified that the constructed matrices in the case of $S^2$ and $T^2$
indeed satisfy the properties seen 
in the covariant derivative interpretation.

In regularizing the covariant derivatives, 
we introduced the embedding coordinates $X^A$ ($A=1,\ldots,n$
with $n\geq 3$). 
Choosing $X^A$ is likely to specify a way of regularization.
It is expected that physical quantities are independent of the choice 
of $X^A$ in the large-$N$ limit.
We leave verification of this as a future work.
Note that as in the covariant derivative interpretation
two-dimensional
curved spacetimes are represented by only the two matrices
$\mathcal{P}_{\pm}$ although it is needed to embed them into
$\mathbb{R}^n$ $(n\geq 3)$. On the other hand,
for instance, the fuzzy sphere is represented by three matrices.
In this sense, our formulation is different from
the representation of fuzzy manifolds in terms of matrices with finite 
size.

In \cite{Ishii:2008ib}, the matrices $T(X^A)$ in Fig. \hspace{-0.1cm}\ref{fig:2} and the limit 
in which $N\rightarrow\infty$ and $\Xi\rightarrow\infty$ with 
$\Xi \ll N$ kept are used to obtain planar $\mathcal{N}=4$ super
Yang-Mills theory on $\mathbb{R}\times S^3$ from the plane
wave matrix model \cite{Berenstein:2002jq} 
as a generalization of the large-$N$ reduction \cite{Eguchi:1982nm}
to curved spaces. The latter is obtained
by dimensional reducing the former to one dimension.
In that case, $S^3$ is constructed from the three matrices 
$T(X^A)$, while it is constructed from $\mathcal{P}_\pm$ and $\mathcal{P}_3$ in our formulation.
However, the commutators between $T(X^A)$ and $T(\varphi)$ do not give the action of $O_{+-}$ on $T(\varphi)$ unlike the action of 
$\mathcal{P}_3$ on $T(\varphi)$. This difference comes from the existence of $T(\partial_\pm X^A)$ in the first term
of the RHS of \eqref{eq:def_Ppm}.

Some future directions are in order.
First, we need to generalize the construction in this paper to higher 
dimensional cases.
It is interesting to perform perturbative calculation of physical 
quantities 
using matrices with finite size
in the covariant derivative interpretation.
Finally, we would like to derive geometries from the results in the 
numerical simulations
of the type IIB matrix model 
by using the formulation developed in this paper.

\section*{Acknowledgments}
We would like to thank Satoshi Kanno, Tatsuya Seko and Harold Steinacker for discussions.
The work of A.T. was supported in part by JSPS KAKENHI Grant Numbers JP21K03532.

\appendix
\renewcommand{\theequation}{A.\arabic{equation}}
\setcounter{equation}{0}
\section{Evaluation of $\mathcal{P}_3$ on $S^2$ and $T^2$}\label{appA}
In this appendix, we evaluate $\mathcal{P}_3$ on $S^2$ and $T^2$.
First, we show that $\mathcal{P}_3$ is given on $S^2$ as \eqref{eq:P3_on_S2}.
By using \eqref{eq:toeplitz_su(2)}, we calculate 
\eqref{eq:P3_N_to_infinity} as follows:
   \begin{align}
     &\big(\mathcal{P}_3T(\varphi)\big)^{(2s',2s)} \notag\\
     &= \hbar_N^{-2} \big(T(\partial_+ X^A)\big)^{(2s',2s'+2)}\big(T(\partial_- X^B)\big)^{(2s'+2,2s')}[[T(X^A),T(X^B)],T(\varphi)]^{(2s',2s)}+\mathcal{O}\qty(\frac{1}{N}) \notag\\
     &= -2i \hbar_N^{-2} \epsilon^{ABC} \big(T(\partial_+ X^A)\big)^{(2s',2s'+2)}\big(T(\partial_- X^B)\big)^{(2s'+2,2s')} \notag\\
     &\quad \times \bigg( \frac{1}{N+2s'+1}\big(T(X^C)\big)^{(2s',2s')}\big(T(\varphi)\big)^{(2s',2s)} - \frac{1}{N+2s+1}\big(T(\varphi)\big)^{(2s',2s)}\big(T(X^C)\big)^{(2s,2s)}\bigg)\notag\\
     &\quad+\mathcal{O}\qty(\frac{1}{N}) \notag\\
     &=\frac{4i(s'-s)}{(N+2s'+1)(N+2s+1)}\hbar_N^{-2}\epsilon^{ABC} \big(T(\partial_+ X^A)\big)^{(2s',2s'+2)}\big(T(\partial_- X^B)\big)^{(2s'+2,2s')}\notag\\
     &\quad \times \big(T(X^C)\big)^{(2s',2s')}\big(T(\varphi)\big)^{(2s',2s)}\notag\\
     &\quad- \frac{2i}{N+2s+1} \hbar_N^{-2} \epsilon^{ABC} \big(T(\partial_+ X^A)\big)^{(2s',2s'+2)}\big(T(\partial_- X^B)\big)^{(2s'+2,2s')}[T(X^C),T(\varphi)]^{(2s',2s)}\notag\\
     &\quad+\mathcal{O}\qty(\frac{1}{N}) \ .
     \label{P_3T}
   \end{align}
We recall that $s$ and $s'$ satisfy $|s|,|s'| \ll N $.
By using the asymptotic properties of 
the Toeplitz operator, 
\eqref{eq:tp_product_N_to_infinity} and \eqref{eq:tp_commutator_N_to_infinity},
we can evaluate the second term in the most right-hand side of \eqref{P_3T} as
    \begin{align}
      & - \frac{2i}{N+2s+1} \hbar_N^{-2} \epsilon^{ABC} \big(T(\partial_+ X^A)\big)^{(2s',2s'+2)}\big(T(\partial_- X^B)\big)^{(2s'+2,2s')}[T(X^C),T(\varphi)]^{(2s',2s)} \notag \\
      & = \frac{N}{N+2s+1}\big(T(\epsilon^{ABC}\partial_+X^A \partial_- X^B \{X^C,\varphi\})\big)^{(2s',2s)} + \mathcal{O}\qty(\frac{1}{N})  \ .
    \end{align}
    Furthermore, noting that the embedding coordinates $X^A$ of $S^2$
    satisfy that $\{X^A,X^B\}=\epsilon^{ABC}X^C$, we obtain
    \begin{align}
      \epsilon^{ABC}\partial_+X^A \partial_- X^B \{X^C,\varphi\} 
      &= \frac{i}{4}\epsilon^{ABC} \{ X^A,X^B\}\{X^C,\varphi\} \notag \\
      &= \frac{i}{4}\epsilon^{ABC} \epsilon^{ABD} X^D \{X^C,\varphi\} \notag \\
      &= X^C ( \partial_+ X^C \partial_- \varphi - \partial_- X^C \partial_+ \varphi) \notag \\
      &= \frac{1}{2} ( \partial_+(X^CX^C) \partial_-\varphi - \partial_-(X^CX^C) \partial_+\varphi)=0 \ .
    \end{align}
    Hence, (\ref{P_3T}) leads to
    \begin{align}
      &\big(\mathcal{P}_3T(\varphi)\big)^{(2s',2s)} \notag \\
      &= \frac{i(s'-s)N^2}{(N+2s'+1)(N+2s+1)}\epsilon^{ABC} \big(T(\partial_+ X^A)\big)^{(2s',2s'+2)}\big(T(\partial_- X^B)\big)^{(2s'+2,2s')} \notag \\ 
      &\times \big(T(X^C)\big)^{(2s',2s')}\big(T(\varphi)\big)^{(2s',2s)} + \mathcal{O}\qty(\frac{1}{N}) \ .
      \label{P_3T2}
    \end{align}
    Finally, substituting the concrete forms of the Toeplitz operators on 
    $S^2$, \eqref{Toeplitz operator of X^A} and
    \eqref{Toeplitz operator of partial X^A}, into \eqref{P_3T2} yields
    \begin{align}
      \big(\mathcal{P}_3T(\varphi)\big)^{(2s',2s)} 
      &= - \frac{s'-s}{2} \frac{N^2(N+2s'-1)}{(N+2s+1)(N+2s'+1)^2}\big(T(\varphi)\big)^{(2s',2s)} + \mathcal{O}\qty(\frac{1}{N}) \notag \\
      &= -\frac{s'-s}{2} \big(T(\varphi)\big)^{(2s',2s)} + \mathcal{O}\qty(\frac{1}{N}) \ .\label{P_3T3}
    \end{align}

The above result can also be obtained by using the asymptotic expansion 
\eqref{eq:aymptotic_expansion_Toeplitz}.
From \eqref{eq:c0}-\eqref{eq:c2}, $[T(X^A),T(X^B)]^{(2s,2s)}$ is evaluated as
\begin{align}
    &[T(X^A),T(X^B)]^{(2s,2s)} \notag\\
    &= \hbar_N \big(T(\{X^A,X^B\})\big)^{(2s,2s)}+ \hbar_N^2\big(T(\widetilde{C}_2(X^A,X^B))\big)^{(2s,2s)}+\hbar_N^{2}\bigg(T\Big(iF^{(E)}_{12}\{X^A,X^B\}\Big)\bigg)^{(2s,2s)}\notag\\
    &\quad +\mathcal{O}\qty(\frac{1}{N^3})\ , \label{eq:asymptotic_expansion_commutator}
\end{align}
where $\widetilde{C}_2(X^A,X^B)$ is defined as
\begin{align}
    \widetilde{C}_2(X^A,X^B) 
    &= C_2(X^A,X^B)-C_2(X^B,X^A)-iF^{(E)}_{12}\{X^A,X^B\}\notag\\
    &= \frac{i}{4}\qty(R\{X^A,X^B\} + (g^{ab}W^{cd}+W^{ab}g^{cd})(\nabla_a\partial_cX^A)(\nabla_b\partial_dX^B))\ . 
\end{align}
The first term in the RHS of 
\eqref{eq:asymptotic_expansion_commutator} contributes to $\mathcal{P}_3T(\varphi)$ in \eqref{eq:P3_N_to_infinity} as
\begin{align}
    &\hbar_N^{-1} \big(T(\partial_+ X^A)\big)^{(2s',2s'+2)}\big(T(\partial_- X^B)\big)^{(2s'+2,2s')}[T(\{X^A,X^B\}),T(\varphi)]^{(2s',2s)}\notag\\
    &=\big(T(\partial_+ X^A \partial_- X^B \{\{X^A,X^B\},\varphi\})\big)^{(2s',2s)}+ \mathcal{O}\qty(\frac{1}{N})\ .
\end{align}
The argument of the Toeplitz operator in the last line is calculates as 
\begin{align}
    \partial_+ X^A \partial_- X^B \{\{X^A,X^B\},\varphi\} 
    &= \epsilon^{ab}\epsilon^{cd} \partial_+ X^A \partial_- X^B\qty(\partial_a\partial_c X^A \partial_d X^B + \partial_c X^A \partial_a\partial_d X^B)\nabla_b \varphi \notag\\
    &= \epsilon^{ab}\epsilon^{-+} \partial_+ X^A \partial_- X^B\qty(\partial_a\partial_- X^A \partial_+ X^B + \partial_- X^A \partial_a\partial_+ X^B)\nabla_b \varphi \notag\\
    &= \epsilon^{ab}\epsilon^{-+} \delta_{+-}\qty(\partial_+ X^A\partial_a\partial_- X^A  + \partial_- X^A \partial_a\partial_+ X^A)\nabla_b \varphi\notag\\
    &= \epsilon^{ab}\epsilon^{-+} \delta_{+-} (\partial_a\delta_{+-})\nabla_b \varphi \notag\\
    &=0\ .
\end{align}
In the second line, we used 
$\partial_\pm X^A \partial_a \partial_\pm X^A =\frac{1}
{2}\partial_a\delta_{\pm\pm}=0$.
Thus, the first term in the 
RHS of \eqref{eq:asymptotic_expansion_commutator} 
does not contribute to $\mathcal{P}_3T(\varphi)$ to $\mathcal{O}(1)$.
Next, we consider the second term 
in the RHS of \eqref{eq:asymptotic_expansion_commutator}.
Substituting this term into \eqref{eq:P3_N_to_infinity}, we obtain its contribution to $\mathcal{P}_3T(\varphi)$ as
\begin{align}
    \big(T(\partial_+ X^A)\big)^{(2s',2s'+2)}\big(T(\partial_- X^B)\big)^{(2s'+2,2s')}[T(\widetilde{C}_2(X^A,X^B)),T(\varphi)]^{(2s',2s)}\ . \label{eq:effect_C'2}
\end{align}
This is evaluated 
as $\mathcal{O}(N^{-1})$ by using the asymptotic behavior 
\eqref{eq:tp_product_N_to_infinity} and 
\eqref{eq:tp_commutator_N_to_infinity}.
The calculations so far hold for the general compact Riemann surfaces.

Finally, by considering the contribution of the third term in the 
RHS of \eqref{eq:asymptotic_expansion_commutator}, we obtain
\begin{align}
    &\big(\mathcal{P}_3T(\varphi)\big)^{(2s',2s')} \notag\\
    &=  \big(T(\partial_+ X^A)\big)^{(2s',2s'+2)}\big(T(\partial_- X^B)\big)^{(2s'+2,2s')} \bigg[\Big(T\Big(iF^{(E')}_{12}\{X^A,X^B\}\Big)\Big)^{(2s',2s')}\big(T(\varphi)\big)^{(2s',2s)}\notag\\
    &\quad -\big(T(\varphi)\big)^{(2s',2s)}\Big(T\Big(iF^{(E)}_{12}\{X^A,X^B\}\Big)\Big)^{(2s,2s)}\bigg]+\mathcal{O}\qty(\frac{1}{N})\ . \label{eq:F_P3T}
\end{align}
Here, we note that in the asymptotic property 
\eqref{eq:tp_commutator_N_to_infinity} the function $f$ is independent 
of the choice of the charge $s$.
Because $F^{(E)}$ depends on the charge $s$ in general, we 
cannot apply \eqref{eq:tp_commutator_N_to_infinity} to
\eqref{eq:F_P3T}. 
We need to consider this contribution which might be $\mathcal{O}(1)$.
For $S^2$, $F^{(E)}_{12}$ is given by $F^{(E)}_{12}=2s\omega_{12}/V=s$, 
hence \eqref{eq:F_P3T} is evaluated as 
\begin{align}
    &\big(\mathcal{P}_3T(\varphi)\big)^{(2s',2s)} \notag\\ 
    &=  \big(T(\partial_+ X^A)\big)^{(2s',2s'+2)}\big(T(\partial_- X^B)\big)^{(2s'+2,2s')} 
    \bigg[\Big(T\Big(iF^{(E')}_{12}\{X^A,X^B\}\Big)\Big)^{(2s',2s')}\big(T
    (\varphi)\big)^{(2s',2s)}\notag\\
    &\quad -\big(T(\varphi)\big)^{(2s',2s)}\Big(T\Big(iF^{(E)}_{12}
    \{X^A,X^B\}\Big)\Big)^{(2s,2s)} \bigg]+\mathcal{O}\qty(\frac{1}{N})\notag\\
    &=i\big(T(\partial_+ X^A)\big)^{(2s',2s'+2)}\big(T(\partial_- X^B)\big)^{(2s'+2,2s')} 
    \bigg[s'\big(T(\{X^A,X^B\})\big)^{(2s',2s')}\big(T(\varphi)\big)^{(2s',2s)}\notag\\
    &\quad -s\big(T(\varphi)\big)^{(2s',2s)}\big(T(
    \{X^A,X^B\})\big)^{(2s,2s)}\bigg]+\mathcal{O}\qty(\frac{1}{N})\notag\\
    &=i(s'-s)\big(T(\partial_+ X^A)\big)^{(2s',2s'+2)}\big(T(\partial_- 
    X^B)\big)^{(2s'+2,2s')} \big(T(\{X^A,X^B\})\big)^{(2s',2s')}\big(T(\varphi)\big)^{(2s',2s)}\notag\\
    &\quad -is\big(T(\partial_+ X^A)\big)^{(2s',2s'+2)}\big(T(\partial_- 
    X^B)\big)^{(2s'+2,2s')} [T(\varphi),T(\{X^A,X^B\})]^{(2s',2s)}+\mathcal{O}\qty(\frac{1}
    {N})\notag\\
    &=i(s'-s)\big(T(\partial_+ X^A\partial_- X^B
    \{X^A,X^B\})\big)^{(2s',2s')}\big(T(\varphi)\big)^{(2s',2s)}+\mathcal{O}\qty(\frac{1}{N})  \ .
    \label{eq:F_P3T_S2}
\end{align}
Furthermore, we can calculate 
$\partial_+ X^A\partial_- X^B\{X^A,X^B\}$ as
\begin{align}
    \partial_+ X^A\partial_- X^B\{X^A,X^B\}
    &=\epsilon^{ab} \partial_+ X^A\partial_- X^B\partial_a X^A\partial_b X^B \notag\\
    &=\epsilon^{-+} \partial_+ X^A\partial_- X^B\partial_- X^A\partial_+ X^B \notag \\
    &=\epsilon^{-+} \delta_{+-}\delta_{+-}\notag \\
    &=\frac{i}{2}\ .
\end{align}
Thus, \eqref{eq:F_P3T_S2} agrees with \eqref{P_3T3}.

In the case of $T^2$, $F^{(E)}_{12}$ is zero, so that $\mathcal{P}_3T(\varphi)$ is evaluated as $\mathcal{O}(N^{-1})$.

\renewcommand{\theequation}{B.\arabic{equation}}
\setcounter{equation}{0}
  \section{Concrete forms of $\mathcal{P}_\pm$ on $S^2$} \label{app:P+-_S2}
   The concrete form of $\mathcal{P}_\pm$ in $S^2$ can be calculated as
    \begin{align}
      &(\mathcal{P}_+)^{(2s'-2,2s)(2s',2s)}_{(mn)(kl)} \notag\\
      &= i \frac{N}{2} \left\{ \frac{N+2s-2n-1}{N+2s+1}\sqrt{\frac{(m+1)(N+2s-m-2)}{(N+2s')(N+2s'-1)}}\delta_{m+1,k}\delta_{n,l}\right.\notag \\
      &\,\,\, -\frac{1}{N+2s+1}\sqrt{\frac{(N+2s'-m-1)(N+2s'-m-2)n(N+2s-n)}{(N+2s')(N+2s'-1)}}\delta_{m,k}\delta_{n-1,l}\notag \\
      &\,\,\, \left. +\frac{1}{N+2s+1}\sqrt{\frac{(m+2)(m+1)(n+1)(N+2s-n-1)}{(N+2s')(N+2s'-1)}}\delta_{m+2,k}\delta_{n+1,l}\right\} \ ,\label{eq:P+_concrete}\\
      &(\mathcal{P}_-)^{(2s'+2,2s)(2s',2s)}_{(mn)(kl)} \notag\\
      &= -i \frac{N}{2} \left\{ \frac{N+2s-2n-1}{N+2s+1}\sqrt{\frac{m(N+2s'-m-1)}{(N+2s'+2)(N+2s'+1)}}\delta_{m-1,k}\delta_{n,l}\right.\notag \\
      &\,\,\, -\frac{1}{N+2s+1}\sqrt{\frac{(N+2s'-m+1)(N+2s'-m)(n+1)(N+2s-n-1)}{(N+2s'+2)(N+2s'+1)}}\delta_{m,k}\delta_{n+1,l}\notag \\
      &\,\,\, \left. +\frac{1}{N+2s+1}\sqrt{\frac{m(m-1)n(N+2s-n)}{(N+2s'+2)(N+2s'+1)}}\delta_{m-2,k}\delta_{n-1,l}\right\} \ .
    \end{align}
   When $n$ and $m$ are regarded as $\mathcal{O}(N)$,
$\mathcal{P}_{\pm}$ behave as $\mathcal{O}(N)$.
    However, $\mathcal{P}_\pm$ are defined in such a way that they act 
    not on general 
    matrices but on the Toeplitz operators that are defined for the 
    fields on $S^2$.
    The $\mathcal{O}(N)$ contributions should be cancelled when 
    $\mathcal{P}_{\pm}$
    operators act on the Toeplitz operators on $S^2$. 
    In order to see this, we consider
    the monopole harmonic functions. The field with any charge can be 
    expressed as a linear combination of the monopole harmonic functions.
    The monnopole harmonic functions with the charge $q$ can be represened in terms of the polar coordinates $(r,\theta,\phi)$ in $\mathbb{R}^3$ as
    \begin{align}
      Y_{j,q}{}^a(\theta , \phi) = N_{j,q}{}^a (1-\cos\theta)^{-\frac{q+a}{2}}(1+\cos\theta)^\frac{q-a}{2}P_{\gamma}{}^{\alpha\beta}(\cos\theta) e^{i(q+a)\phi} \ ,
    \end{align}
    where $j=|q| ,|q|+1,\dots$, $a=-j,-j+1,\dots,j$,
    $\alpha=-q-a$, $\beta=q-a$ and $\gamma=j+a$. 
    $N_{j,q}{}^a$ is the normalization factor, and 
    $P_\gamma{}^{\alpha\beta}$ is the Jacobi polynomials defined by
    \begin{equation}\label{eq:jacobi_polynomial}
      P_\gamma{}^{\alpha\beta} (x)= \frac{(-1)^\gamma}{2^\gamma n!} (1-x)^{-\alpha}(1+x)^{-\beta} \frac{d^\gamma}{dx^\gamma}[(1-x)^\alpha(1+x)^\beta(1-x^2)^\gamma] \ .
    \end{equation}
    For $q \ge 0$, by performing the derivatives in \eqref{eq:jacobi_polynomial}
    and using the stereographic projection $(z,\bar{z})$, we obtain
    \begin{align}
      Y_{j,q}{}^a (z,\bar{z}) =& \sum_{b=0}^{\gamma+\alpha} \sum_{c=0}^{\beta-\alpha} \sum_{d=0}^{2b+c-\gamma} N_{j,q}{}^a \frac{(-1)^{\gamma+b+d}(2b+c)!}{2^{\gamma-a}\gamma !(2b+c-\gamma)!}\binom{\beta-\alpha}{c}\binom{\gamma+\alpha}{b}\binom{2b+c-\gamma}{d}\notag\\ 
      &\times\frac{z^{d+q+a}\bar{z}^d}{(1+|z|^2)^{2b+c-j}} \ ,
      \label{monopole harmonic function}
    \end{align}
   
We extract functions 
    \begin{equation}
      f^{(q)}(z,\bar{z}) = \frac{z^{b+q+a}\bar{z}^b}{(1+|z|^2)^c} 
    \end{equation}
from \eqref{monopole harmonic function} where we made the replacement $d\rightarrow b$ and $2b+c-j \rightarrow c$, and calculate the corresponding Toeplitz operators.
We evaluate the large-$N$ behavior of $\mathcal{P}_+$ in \eqref{eq:P+_concrete} when $\mathcal{P}_+$ is acted
on these Toeplitz operators.
For simplicity, we consider a case in which the charge $s$ of 
the fiber bundle $E$ is zero.
    The Toeplitz operators are given by
    \begin{align}
      \big(T(f^{(q)})\big)_{mn} =& \delta_{n+a+q,m} 
      \sqrt{\frac{N!(N+2q)!}{m!(N+2q-m-1)!(m-a-q)!(N-m+a+q-1)!}}\notag\\
      &\times\frac{(m+b)!(N+c+q-m-b-1)!}{(N+c+q)!} \ .
      \label{eq:Toeplitz_of_monopole}
    \end{align}
    By using the Stirling formula 
    \begin{equation}
      n! \simeq \sqrt{2\pi n} \qty(\frac{n}{e})^n \ ,
    \end{equation}
    we see that 
     the RHS of \eqref{eq:Toeplitz_of_monopole}
     behaves as $\mathcal{O}(1)$. Here we note that
     we regard $m$ and $(N-m)$ as $\mathcal{O}(N)$.
    Next, we calculate the action of $\mathcal{P}_+$ as
    \begin{align}
      \big(\mathcal{P}_+T(f^{(q)})\big)_{mn} =& i\delta_{n+j+q,m+1} \frac{N}{N+1}\frac{(m+b)!(N+c+q-m-b-3)!}{(N+c+q)!}\notag\\
      &\times \sqrt{\frac{N!(N+2q-2)!}{m!(N+2q-m-3)!(m-j-q+1)!(N-m+j+q-2)!}}\notag\\
      &\times \{(N-2m+2j+2q-2)(m+b+1)(N+c+q-m-b-2)\notag\\
      &-(m-j+q+1)(N+c+q-m-b-1)(N+c+q-m-b-2)\notag\\
      &+(N-m+j+q-2)(m+b+2)(m+b+1)\}\label{eq:P+Tf} \ .
    \end{align}
    We see that the curly bracket in the RHS of \eqref{eq:P+Tf}
    behaves as $\mathcal{O}(N^2)$, while the factor including factorials
    is estimated as $\mathcal{O}(N^{-2})$ by using the Stirling formula.
    Thus, we have verified that the RHS of \eqref{eq:P+Tf}
    behaves as $\mathcal{O}(1)$.
    This is also true for $\mathcal{P}_-$.
\bibliography{reference}
\bibliographystyle{utphys}
\end{document}